# Engineering spin-orbit coupling for photons and polaritons in microstructures

V. G. Sala[1,2], D. D. Solnyshkov[3], I. Carusotto[4], T. Jacqmin[1], A. Lemaître[1], H. Terças[3], A. Nalitov[3], M. Abbarchi[1,5], E. Galopin[1], I. Sagnes[1], J. Bloch[1], G. Malpuech[3], A. Amo[1]

[1]*Laboratoire de Photonique et Nanostructures, LPN/CNRS, Route de Nozay, 91460 Marcoussis, France*

[2]*Laboratoire Kastler Brossel, Université Pierre et Marie Curie, École Normale Supérieure et CNRS, UPMC Case 74, 4 place Jussieu, 75252 Paris Cedex 05, France*

[3]*Institut Pascal, PHOTON-N2, Clermont Université, University Blaise Pascal, CNRS, 24 avenue des Landais, 63177 Aubière Cedex, France*

[4]*INO-CNR BEC Center and Dipartimento di Fisica, Università di Trento, I-38123, Povo, Italy*

[5]*Laboratoire Pierre Aigrain, École Normale Supérieure, CNRS (UMR 8551), Université Pierre et Marie Curie, Université D. Diderot, 75231 Paris Cedex 05, France*

**One of the most fundamental properties of electromagnetism and special relativity is the coupling between the spin of an electron and its orbital motion[1]. This is at the origin of the fine structure in atoms, the spin Hall effect in semiconductors[2], and underlies many intriguing properties of topological insulators, in particular their chiral edge states[3,4]. Configurations where neutral particles experience an effective spin-orbit coupling have been recently proposed and realized using ultracold atoms[5,6] and photons[7-10]. Here we use coupled micropillars etched out of a semiconductor microcavity to engineer a spin-orbit Hamiltonian for photons and polaritons in a microstructure. The coupling between the spin and orbital momentum arises from the polarisation dependent confinement and tunnelling of photons between micropillars arranged in the form of a hexagonal photonic molecule. Dramatic consequences of the spin-orbit coupling are experimentally observed in these structures in the wavefunction of polariton condensates, whose helical shape is directly visible in the spatially resolved polarisation patterns of the emitted light. The strong optical nonlinearity of polariton systems suggests exciting perspectives for using quantum fluids of polaritons[11] for quantum simulation of the interplay between interactions and spin-orbit coupling.**

Spin-orbit (SO) coupling is the coupling between the momentum and spin of a particle. It gives rise to the fine structure in atomic spectra and it is present in some bulk materials. A prominent example is semiconductors without inversion symmetry, in which static electric fields present in the crystal Lorentz transform to a magnetic field in the reference frame of a moving electron, which then couples to the electron spin. The resulting SO coupling leads to a number of exciting phenomena like the spin Hall effect[2,12], the persistent spin helix[13], or topological insulation in the absence of any external magnetic field[3].

In semiconductors the SO coupling is determined by the crystalline structure being, therefore, hard to manipulate and often difficult to separate from other effects. Novel systems, like ultracold atomic gases under suitably designed optical and/or magnetic field configurations[6,14], and photons in properly designed structures[15], allow for a great flexibility and control of the system Hamiltonian. Even though particles without magnetic moment, such as photons, cannot experience the usual SO coupling, the engineering of an effective Hamiltonian acting on photons in structured media has led, for instance to the observation of the photonic analogue of the spin Hall effect in planar structures[16-18], and unidirectional photon transport in lattices with topological protection from disorder scattering[7,19-21]. Effective SO couplings have been induced in arrays of photonic ring resonators using the spin-like degree of freedom associated to the rotation direction of photons in the resonator[8,21]. A promising perspective is to use the intrinsic photon spin: the polarisation degree of freedom[10]. In combination with the strong spin-dependent interactions naturally present in microcavity-polariton devices and the possibility of scaling up to lattices of arbitrary geometry[22-24], the realization of such a coupling will open the way to the simulation of many-body effects in a new quantum optical context[11]. Some envisioned examples would be the controlled nucleation of fractional topological excitations[25,26], the self-organisation of polarisation patterns[27], the simulation of spin models using photons[28], or the generation of fractional quantum Hall states for light[9].

In this Letter we engineer the coupling between the polarisation (spin) and the momentum (orbital) degrees of freedom for photons and polaritons using a photonic microstructure with a ring-like shape. The structure is a hexagonal chain of overlapping micropillars as shown in Figure 2a. Each individual micropillar shows discrete confined photonic modes, the lowest one with cylindrical $s$ symmetry. Thanks to the spatial overlap of adjacent micropillars the confined photons can tunnel between neighbouring sites[29,30]. As it was shown in Ref. 31, the tunnelling amplitude through a given link is different for the polarisation states parallel and orthogonal to the link direction (Fig. 1a,b). We show here that when extended to the hexagonal structure, this polarisation dependent tunnelling together with on-site energy splittings, results in an effective SO coupling for photons. In our system, photons are strongly coupled to quantum-well excitons giving rise to polariton states, which hold the same spin/polarisation properties as the confined photons. We show that the engineered SO coupling drives the Bose-Einstein condensation of polaritons into states with complex spin textures.

Within a tight-binding formalism, the system can be modelled along the lines sketched in Fig. 1. For each link connecting the pillars $j \leftrightarrow j+1$, we can define a pair of (real) unit vectors $e_L^{(i)}$ and $e_T^{(i)}$, respectively longitudinal and transverse to the link direction (see Fig. 1c). In the absence of particle-particle interactions, the tight-binding Hamiltonian describing the six coupled pillars reads:

$$H = -\sum_{j=1}^{6}\left\{\left[\hbar t_L\left(\hat{a}_{j+1}^{\dagger}\cdot e_L^{(j)}\right)\left(e_L^{(j)\dagger}\cdot \hat{a}_j\right) + \hbar t_T\left(\hat{a}_{j+1}^{\dagger}\cdot e_T^{(j)}\right)\left(e_T^{(j)\dagger}\cdot \hat{a}_j\right)\right] + h.c. + \right.$$
$$\left. + \frac{\Delta E}{3}\left[\left(\hat{a}_j^{\dagger}\cdot\left(e_L^{(j)}+e_L^{(j-1)}\right)\right)\left(\left(e_L^{(j)\dagger}+e_L^{(j-1)\dagger}\right)\cdot\hat{a}_j\right) - \left(\hat{a}_j^{\dagger}\cdot\left(e_T^{(j)}+e_T^{(j-1)}\right)\right)\left(\left(e_T^{(j)\dagger}+e_T^{(j-1)\dagger}\right)\cdot\hat{a}_j\right)\right]\right\},$$
(1)

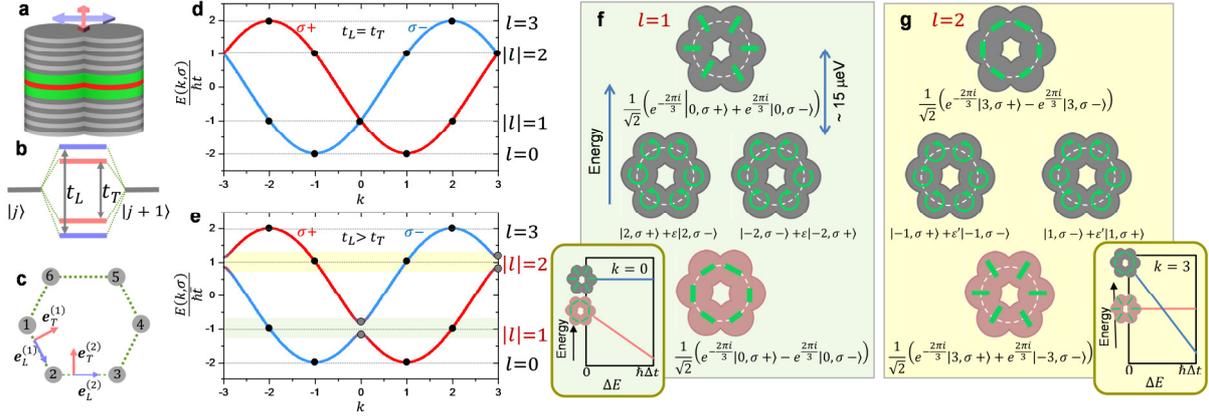

Fig. 1. **Spin-orbit split states**. **a** Scheme of two coupled pillars showing different tunnelling probabilities for photons polarised longitudinal (light blue) and transverse (pink) to the link. **b** This results in polarisation splitting of the bonding and anti-bonding states. **c** Scheme of the unit vectors perpendicular and transverse to each link used to calculate the SO-coupled eigenstates of Hamiltonian (1). **d** Polarisation dependent classification of the energy states in the absence of SO coupling ($t_L = t_T$ and $\Delta E = 0$) as a function of their total momentum $k = l + \sigma$. States located on each horizontal line have a well-defined orbital momentum $l$ and form multiplets. **e** Same as **d** for $t_L \gtrsim t_T$. The SO coupling is evidenced by the anticrossing of the bands, at $k = 0$ and $k = 3$. **f**, **g**, Fine structure and polarisation pattern of the molecular eigenstates corresponding to the multiplets $l = 1$, **f**, and $l = 2$, **g**, split by the SO coupling for $\Delta t = 30\mu eV$, $\Delta E = 0$. Note that $\varepsilon, \varepsilon' \ll 1$. The salmon coloured molecules show the states in which condensation takes place in Figs. 3 and 4. The insets in **f**,**g** show the relative energy of the top and bottom states of each multiplet when $\Delta E > 0$.

where $\hat{a}_j = \boldsymbol{e}_{\sigma+}\hat{a}_{j,\sigma+} + \boldsymbol{e}_{\sigma-}\hat{a}_{j,\sigma-}$ is the vector field operator for polaritons. For each site $j$ on the ring, the $\hat{a}_{j,\sigma+(\sigma-)}$ operators destroy a polariton in the $\sigma+ (\sigma-)$ circular polarisation basis. $t_L$ and $t_T$ are the tunnelling amplitudes for photons with linear polarisation oriented along and transverse to the link direction, respectively (see Fig. 1b). Finally, the $\Delta E$ terms provide an on-site splitting between linearly polarised states oriented azimuthally and radially to the hexagon. These terms account for the waveguide-like geometry of the molecule[32]: if instead of a hexagonal chain we would have a uniform ring guide, these terms would be the dominant contribution to the SO coupling effects. Note that in Hamiltonian (1) the rigid rotation of the unit vectors $\boldsymbol{e}_L^{(j)}$, $\boldsymbol{e}_T^{(j)}$ while going around the hexagon will be crucial to describe the SO coupling.

In the case $\Delta t \equiv t_L - t_T = 0$ and $\Delta E = 0$, the spin and the momentum are decoupled and Hamiltonian (1) reduces to $H = -\sum_{j=1}^{6} \frac{\hbar t}{2}\left(\hat{a}_{j+1}^\dagger \cdot \hat{a}_j + \hat{a}_j^\dagger \cdot \hat{a}_{j+1}\right)$. This is analogous to the Hamiltonian that describes the tunnelling between electronic $\pi$ states of the $C_6H_6$ benzene molecule. The eigenfunctions are delocalised over the six micropillars and can be classified in terms of the orbital angular momentum $l = 0, \pm 1, \pm 2, 3$, which determines the relative phase of the lobe of each micropillar (Fig. 2c): $\hat{a}_{k,\sigma} = \sum_{j=1}^{6} 6^{-1/2} e^{i\pi l j/3} \hat{a}_{j,\sigma}$. The wavefunction of the $l = 0$ state presents a constant phase over all the pillars, while $l = \pm 1, \pm 2$ states contain phase vortices of topological charge $l$ (the phase changes by $2\pi l$ when going around the molecule). Finally, the $l = 3$ state presents a phase change of $\pi$ from pillar to pillar. The eigenenergies depend on $l$ as follows: $E(l, \sigma) = -2\hbar t \cos(2\pi l/6)$.

In the general case, when $t_L - t_T \neq 0$ and/or $\Delta E \neq 0$, spin and orbital degrees of freedom are coupled. As it happens to electrons in atoms, in the presence of a finite SO coupling neither the spin nor the orbital angular momentum are good quantum numbers: Hamiltonian (1) is no longer symmetric under separate orbital or spin rotations. Nevertheless, combined spin and orbital rotations remain symmetry elements, and the corresponding conserved quantity is the total angular momentum $k = l + \sigma$. The eigenstates are then better labelled in terms of $k$, which remains a good quantum number. Figure 1d,e shows the dispersion of the eigenstates as a function of $k$ for negligible (Fig. 1d) and significant values of $t_L - t_T > 0$ (Fig. 1d) as predicted by Hamiltonian (1), which can be rewritten in $k$-space as:

$$H \cong -\sum_{k,\sigma}[2\hbar t \cos(2\pi[k-\sigma]/6)\, \hat{a}^\dagger_{k,\sigma}\hat{a}_{k,\sigma}] - \sum_k \left(\frac{\hbar \Delta t}{2}\cos\left(\frac{\pi k}{3}\right) + \frac{\Delta E}{2}\right)[\hat{a}^\dagger_{k,\sigma-}\hat{a}_{k,\sigma+} + h.c.]. \quad (2)$$

As compared to the $l$-dependent dispersion, the $k$-dependent dispersions for the two $\sigma \pm$ spin states are shifted by $\pm 1$ units of angular momentum. The crossing that is visible in Fig. 1d at $k = 0$ and $k = 3$ for $t_L - t_T = 0$ and $\Delta E = 0$, is lifted by the mixing of the two spin components by the SO coupling when $t_L - t_T \neq 0$ and/or $\Delta E \neq 0$ (Fig. 1e), giving rise to a fine structure in the energy spectrum. At $k = 0$ (respectively $k = 3$), the magnitude of the splitting between the two states is $\Delta E_{SO} = |+\hbar \Delta t + \Delta E|$ (respectively $\Delta E_{SO} = |-\hbar \Delta t + \Delta E|$).

With respect to the orbital form, the two $k = 0$ eigenstates are symmetric/antisymmetric combinations of states with opposite circular polarisation and with opposite orbital angular momentum $l = \pm 1$: $\psi(k = 0, \pm) = 2^{-1/2}\left(e^{-\frac{2\pi i}{3}}|k = 0\, (l = -1), \sigma+\rangle \pm e^{\frac{2\pi i}{3}}|k = 0\, (l = 1), \sigma-\rangle\right)$. In real space, the lower (upper) state corresponds to a polarisation texture oriented in the azimuthal (radial) directions, respectively, as represented in Fig. 1f. Note that the two remaining states with $|l| = 1$ ($|l = +1, \sigma+\rangle$, and $|l = -1, \sigma-\rangle$) are far away in energy from their "partner" of same k and opposite spin. They are thus practically unaffected by the SO coupling, and they remain located in between the split apart states.

An analogue situation takes place at $k = 3$: the resulting eigenstates are symmetric and antisymmetric combinations of orbital states with opposite circular polarisation and opposite angular momentum momentum $l = \pm 2$: $\psi(k = 3, \pm) = 2^{-1/2}\left(e^{-\frac{2\pi i}{3}}|k = 3\, (l = 2), \sigma+\rangle \mp e^{\frac{2\pi i}{3}}|k = 3\, (l = -2), \sigma-\rangle\right)$. For small $\Delta E$ ($< \hbar \Delta t$), the ordering in energy is exchanged with respect to the $k = 0$ doublet, giving an azimuthally (respectively, radially) polarised lower (respectively upper) state. As sketched in the inset of Fig. 1g, for increasing positive $\Delta E$, the SO-splitting can be cancelled ($\Delta E = \hbar \Delta t$) and its sign reversed ($\Delta E > \hbar \Delta t$). Note that an alternative description of the SO coupling in terms of an effective magnetic field acting on the pseudo-spin of the photon is given in the Methods and Supplementary Material together with a detailed solution of Hamiltonian (1).

In order to experimentally evidence this engineered SO coupling we use a polaritonic structure fabricated from an AlGaAs $\lambda/2$ microcavity with 12 GaAs quantum wells, a Rabi splitting of 15 meV, and a polariton lifetime of 50 ps (see Methods). We engineer hexagonal molecules made out of six overlapping round micropillars (Fig. 2a). Each micropillar provides a quasi-cylindrical 0D confinement and the overlap results in a first neighbour tunnel coupling[29] $t_L \gtrsim t_T = 0.3$ meV. According to the measurements reported in Ref. 31 for two coupled micropillars, the polarisation-dependent tunnelling term dominates over the on-site splitting ($\Delta t \sim 30\ \mu eV > |\Delta E|$), which we will neglect in the following. The sample is kept at 10 K and excited out of resonance with a Ti:Sapph cw laser, focused in a 12 μm diameter spot, entirely covering a single molecule. The energy resolved photoluminescence is recorded by a CCD camera and its polarisation is analysed in the six Stokes polarisation components: linear vertical (V), horizontal (H), diagonal (D) and antidiagonal (A), and circular $\sigma+$, and $\sigma-$ (see Methods).

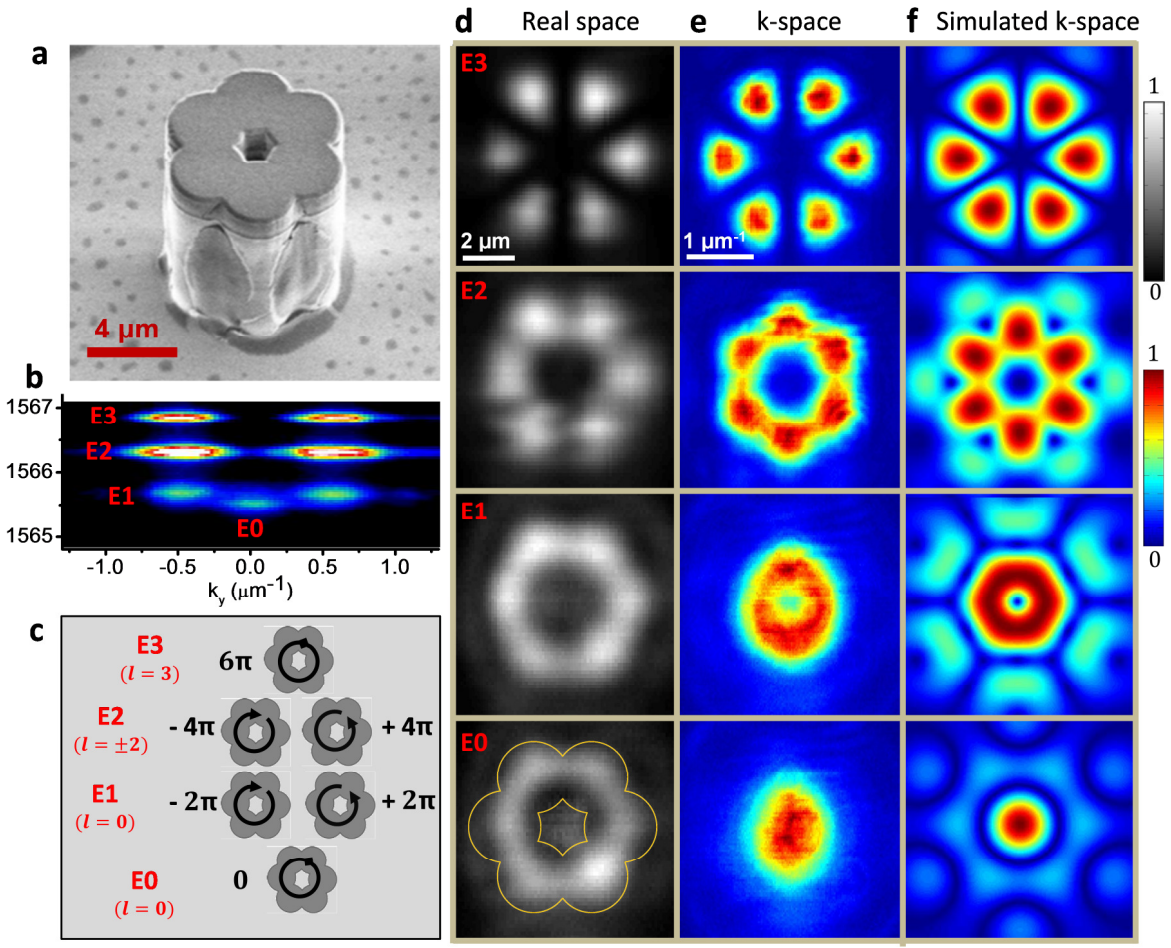

Fig. 2. **The photonic molecular states**. **a** Scanning electron microscope image of the polaritonic molecule. **b** Emission spectrum in momentum space at $k_x = 0.2\ \mu m^{-1}$ and low excitation power (1.5 mW), showing the four molecular states ($E0, ..., E3$) arising from the coupling of the lowest energy mode of each single micropillar. In this low excitation power regime, the spin-orbit coupling is small as compared to the emission linewidth and cannot be resolved. **c** Scheme of the phase winding of the eigenfunctions of the $E(l)$ states without accounting for the spin. **d**, **e** Measured real space, **d**, and momentum space, **e**, emission. **f** Simulation of the momentum space emission obtained by the Fourier transform of the tight binding model eigenfunctions with a Gaussian distribution over each lobe. Solid lines in **d** show the contours of the structure.

At low excitation density, the incoherent relaxation of carriers in the structure populates all the molecular levels (Fig. 2b). As the natural polariton linewidth of 100 μeV is much larger than the polarisation dependent tunnelling $\Delta t$, SO coupling effects are not revealed and the spontaneous emission is effectively unpolarised. The eigenstates can be described as the spinless eigenstates defined above when $t_L = t_T$ and $\Delta E = 0$, with four energy levels (Fig. 2b) and the orbital structure depicted in Fig. 2c. Each molecular eigenmode has six lobes centred on each micropillar, as evidenced in the energy resolved real-space images in Fig. 2d. The orbital phase structure gives rise to distinct patterns in momentum space as shown in Fig. 2e,f and can be directly imaged by performing interferometric measurements[33] (see Supplementary Information).

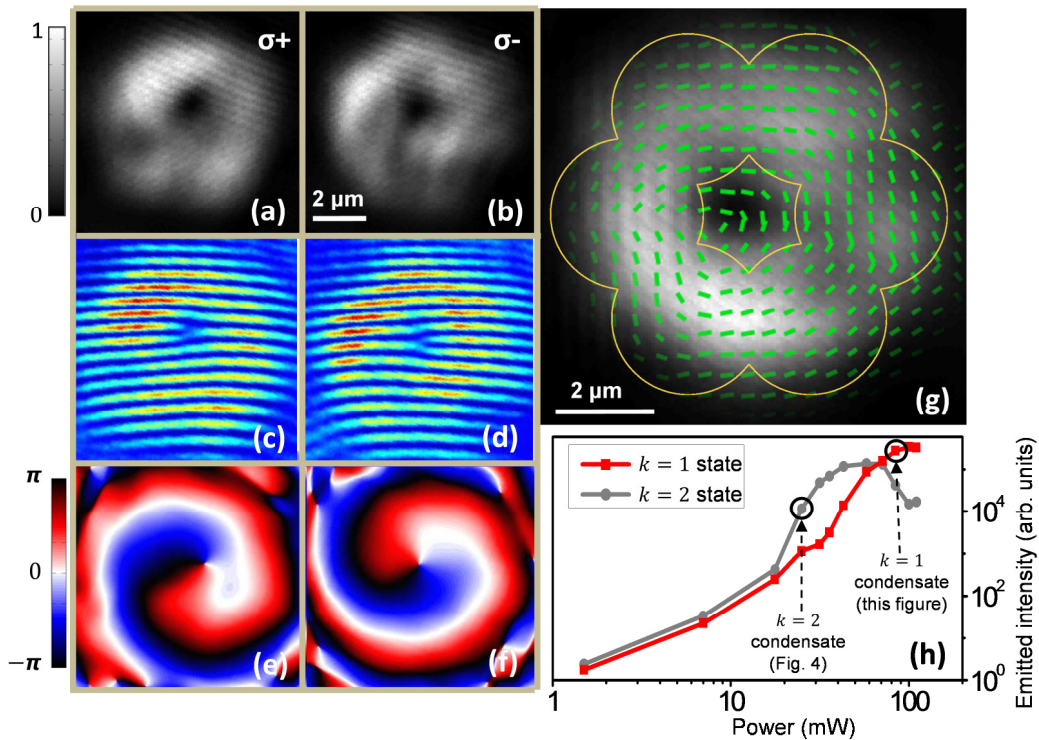

Fig. 3. **Condensation in $k = 0$ state**. **a**, **b** Real space emission in the $\sigma +$ and $\sigma -$ circular polarisations for a polariton condensate in the state $2^{-1/2}\left(e^{-\frac{2\pi i}{3}}|0, \sigma +\rangle - e^{\frac{2\pi i}{3}}|0, \sigma -\rangle\right)$, observed at a pumping intensity of 84 mW. Each polarisation component contains a vortical current in the clockwise ($\sigma +$) and counterclockwise ($\sigma -$) direction, evidenced in the fork-like dislocations apparent in the interferometric measurements shown in **c** and **d**, respectively, and in the extracted phase gradients in **e** and **f**. **g** The green traces show the plane of linear polarisation of the emission measured locally, superimposed to the total emitted intensity of the molecule. The condensate shows azimuthal linear polarisation. Solid lines depict the contour of the structure. **h** Power dependence of the emitted intensity at the energy of the state described in **a-g**, and at the energy of the state described in Fig. 4 (grey points).

At high excitation density, polariton condensation takes place and the population ends up concentrating in a single quantum state with a reduced emission linewidth[34]. As usual in non-equilibrium systems, condensation does not necessarily occur in the ground state and the steady state is determined by a complex interplay between pumping, relaxation and decay (see Supplementary Information). In our structure, in particular, we observe two non-linear thresholds in the emission intensity as a function of pumping intensity, which correspond to condensation in two different states (Fig. 3h). The existence of two thresholds arises from the non-linear power dependence of the polariton relaxation channels[35]. The combination of a reduced linewidth and condensation in excited states grants access to the fine structure caused by SO coupling in the polaritonic benzene molecule. In the case shown in Fig. 3 (excitation intensity of 84 mW), polaritons condense in the lowest state arising from the $|l| = 1$ quadruplet, that is, into the lowest state at $k = 0$ (Fig. 1f). This fact is evidenced when mapping the linear polarisation of the emission (Fig.3g): the polarization is directed in the azimuthal direction around the molecule as predicted in Fig. 1f. The polarisation-selective interferometric analysis of the emission in the circular basis shown in Fig. 3c-f (see Methods) reveals the underlying helical orbital structure of the state, consisting of the linear superposition of two states of opposite orbital vorticity $l = \pm 1$ and opposite spins. For each spin state $\sigma \mp$, the phase winds by $\pm 2\pi$ while looping around the molecule.

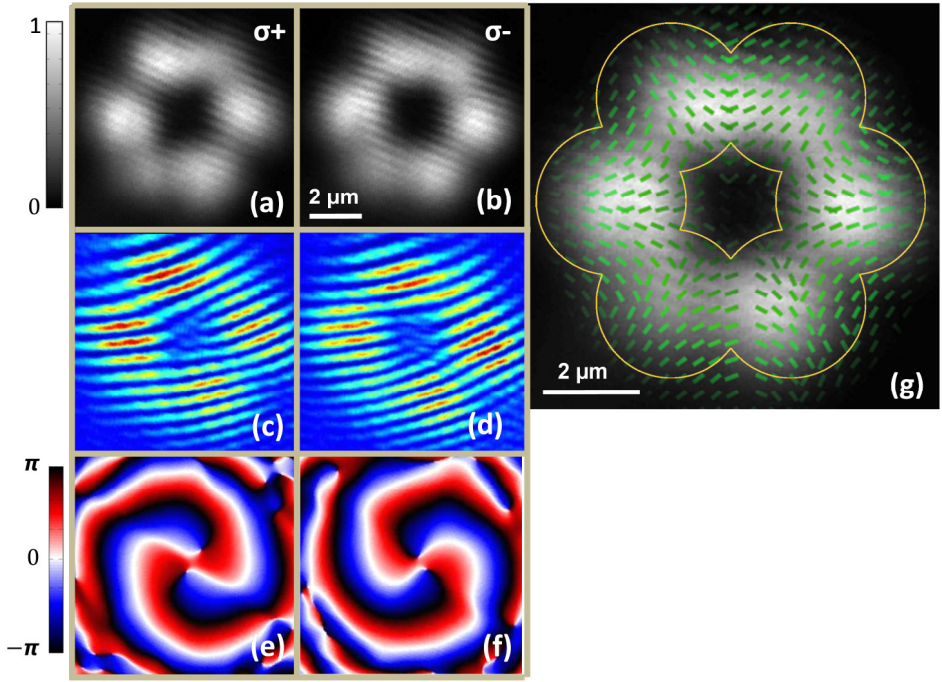

Fig. 4. **Condensation in $k = 1$ state. a** and **b** Real space emission in the $\sigma +$ and $\sigma -$ circular polarisations for a polariton condensate in the state $2^{-1/2}\left(e^{-\frac{2\pi i}{3}}|3, \sigma +\rangle + e^{\frac{2\pi i}{3}}|3, \sigma -\rangle\right)$ observed at 25 mW. Each polarisation component contains a vortical current of double phase charge in the counterclockwise ($\sigma +$) and clockwise ($\sigma -$) directions, evidenced in the double fork-like dislocations apparent in the interferometric measurements shown in **c** and **d**, respectively, and in the extracted phase gradients in **e** and **f**. **g** The green traces show the plane of linear polarisation of the emission measured locally, superimposed to the total emitted intensity of the molecule. The condensate shows a linear polarisation pattern pointing radially in the direction of the micropillars. Solid lines depict the contour of the structure.

Condensation in other SO coupled states can be observed by varying the excitation conditions (Fig. 3h). For a 25 mW excitation density, polariton condensation takes place in the lowest energy state of the $|l|=2$ quadruplet. Within the pillars, the emission is linearly polarised along the radial direction (Fig.4g) as predicted in the lowest state of Fig.1g. On the other hand, the polarisation-selective interferometric images of Fig.4c-f evidence the underlying orbital structure of the state, consisting in the linear superposition of two states of opposite spin and opposite orbital angular momentum. In contrast to the case of Fig. 3, the extracted phase (Fig. 4e,f) changes from zero to $\pm 4\pi$ while looping around the molecule.

In this experiment, polariton condensation occurs selectively into two particular states with the polarisation structures shown in Figs. 3 and 4, evidencing the SO coupling we have engineered for polaritons. The polarisation pattern of the other states should also be accessible by resorting to a resonant excitation scheme as opposed to the non-resonant one used here. The SO coupling reported here for polaritons originates in the polarisation dependence of the photonic confinement and of the photon tunnelling amplitude, which can both be engineered with a suitable design of the structure. For instance, by asymmetrising the micropillars we can enhance $\Delta E$ such that the SO coupling is cancelled in the multiplet $|l|=2$, or its sign reversed, as sketched in the inset of Fig. 2g. Notice that the same SO coupling engineering could be implemented for pure photons, either choosing a larger exciton-photon detuning or processing an empty cavity.

Further promising developments are expected to occur when the SO coupling is scaled up to larger systems such as two-dimensional lattices, where photonic quantum spin Hall states and spin topological insulators[3] can be realized. Exciting new features are anticipated in systems with a high degree of phase frustration, like the optically accessible flat bands recently reported in a honeycomb lattice of micropillars[24]. At strong light intensities, our system appears to be an excellent platform to study the effect of SO coupling on non-linear topological excitations like vortex solitons[36] and non-linear ring states[37]. When the strong polariton nonlinearities are brought towards the single-particle level, new quantum features are expected to originate while polaritons enter a strongly correlated state[11].

**Methods**

**Sample and experimental set-up**. The microcavity sample is grown by Molecular Beam Epitaxy and it is made out of Bragg reflectors with 29 and 40 pairs of alternating $Al_{0.95}Ga_{0.01}As/Al_{0.20}Ga_{0.80}As$ $\lambda/4$ layers that define a $\lambda/2$ cavity ($\lambda$=787nm). Three sets of four GaAs quantum wells of 70 Å in width are distributed at the three central maxima of the confined electromagnetic field. The coupling between the quantum well *1s* excitons and the fundamental longitudinal cavity mode results in a Rabi splitting of 15 meV at 10 K.

The planar structure is dry etched into a hexagonal photonic molecule made out of six coupled micropillars. The diameter of each micropillar is 3 µm, and the centre-to-centre separation is 2.4 µm, resulting in a photonic nearest neighbours coupling[29] of 0.3 meV. The homogeneity of the emitted intensity over all the micropillars seen in Fig. 2d demonstrates negligible defects in the structure. We work at zero photon-exciton detuning, where we measure a polariton lifetime of about 50 ps, corresponding to a photon quality factor of 66000.

Excitation is performed with a cw monomode laser at a wavelength of 735 nm, above the first reflectivity minimum of the stop band defined by the Bragg mirrors. The laser is focused in a spot of 12 µm in diameter pumping the whole molecule. The photoluminescence is collected with a microscope objective and it is energy resolved in a 0.5 m spectrometer attached to a CCD camera. We use a long-wavelength pass filter to remove the stray light from the laser. A set of λ/4 and λ/2 waveplates and linear polarisers allows us selecting the different polarisation components of the emission. With this information we can reconstruct the direction of the linear polarisation of the emission at each point of the molecule (see Supplementary Information).

The relative phase of the emission coming from different micropillars is measured using a modified Michelson interferometer. The real space emission of the molecule interferes in the CCD camera at zero time delay with a magnified image coming from only one of the micropillars, providing a homogeneous phase reference. The two images reach the CCD at different angles, resulting in the interference patterns shown in Fig. 3c,d and Fig. 4 c,d. We extract the spatial phase distribution (shown in Fig. 3e,f and Fig.4e,f) from a Fourier transform analysis[33].

**Spin-Orbit Hamiltonian in operator and effective field forms**

Hamiltonians (1) and (2) can be expressed in the form of an operator acting on a spinor $\vec{\psi}(j) = [\psi^+(j), \psi^-(j)]^T$, with $j = 1, \ldots, 6$ the micropillar index and +,- the two elements of the circular polarisation basis. For this purpose we introduce the diagonal part of the spinor Hamiltonian $\hat{H}_0 = \hat{H}(\Delta t = 0, \Delta E = 0)$. Eigenstates of $\hat{H}_0$ are described by the quantum number $l$:

$$E(l, \sigma) = -2\hbar t \, cos(2\pi l/6)$$

We can introduce an operator $\hat{M} = \frac{\partial^2 E}{\partial l^2} = cos(2\pi l/6)$ allowing us to write the complete Hamiltonian:

$$\hat{H} = \hat{H}_0 - \frac{\Delta E}{2}\begin{pmatrix} 0 & e^{-2i\varphi_j} \\ e^{2i\varphi_j} & 0 \end{pmatrix} - \frac{\hbar \Delta t}{2}\left[\hat{M}\begin{pmatrix} 0 & e^{-2i\varphi_j} \\ e^{2i\varphi_j} & 0 \end{pmatrix} + \begin{pmatrix} 0 & e^{-2i\varphi_j} \\ e^{2i\varphi_j} & 0 \end{pmatrix}\hat{M}\right],$$

where $\varphi_j = j\pi/3$.

In the context of SO coupling in semiconductors, it is meaningful to express the Hamiltonian in terms of an effective magnetic field acting on the spin of the particles. In the case of Hamiltonian (1)-(2), this can be done in cylindrical coordinates the following way:

$$\epsilon \vec{\psi} = -\frac{\hbar^2}{2mR^2}\frac{\partial^2}{\partial \varphi^2}\vec{\psi} - \vec{\Omega} \cdot \vec{\psi}$$

where R is the mean radius of the molecule and $\vec{\Omega}$ is the effective field:

$$\frac{\hbar\vec{\Omega}}{2} = \left(\frac{\hbar\,\Delta t}{2}cos\left(\frac{\pi k}{3}\right) + \frac{\Delta E}{2}\right)\begin{pmatrix} 0 & e^{-2i\varphi_j} \\ e^{2i\varphi_j} & 0 \end{pmatrix}.$$

The polarisation patterns in the eigenmodes can then be understood as those arising from the alignment and antialignement of the photon pseudospin with respect to the effective field $\vec{\Omega}$.

**Acknowledgements**

We thank J. W. Fleischer, G. Molina-Terriza, A. Poddubny, P. Voisin and M. Wouters for fruitful discussions. This work was supported by the French RENATECH, the ANR-11-BS10-001 contract "QUANDYDE", the ANR-11-LABX-0014 Ganex, the RTRA Triangle de la Physique (contract "Boseflow1D"), the FP7 ITN "Clermont4" (235114), the FP7 IRSES "Polaphen" (246912), the POLATOM ESF Network, the Nanosaclay Labex and the ERC grants Honeypol and QGBE.

# Supplementary Information for "Engineering spin-orbit coupling for photons and polaritons in microstructures"

V. G. Sala[1,2], D. D. Solnyshkov[3], I. Carusotto[4], T. Jacqmin[1], A. Lemaître[1], H. Terças[3], A. Nalitov[3], M. Abbarchi[1,5], E. Galopin[1], I. Sagnes[1], J. Bloch[1], G. Malpuech[3], A. Amo[1]

[1]Laboratoire de Photonique et Nanostructures, LPN/CNRS, Route de Nozay, 91460 Marcoussis, France

[2]Laboratoire Kastler Brossel, Université Pierre et Marie Curie, École Normale Supérieure et CNRS, UPMC Case 74, 4 place Jussieu, 75252 Paris Cedex 05, France

[3]Institut Pascal, PHOTON-N2, Clermont Université, University Blaise Pascal, CNRS, 24 avenue des Landais, 63177 Aubière Cedex, France

[4]INO-CNR BEC Center and Dipartimento di Fisica, Università di Trento, I-38123, Povo, Italy

[5]Laboratoire Pierre Aigrain, École Normale Supérieure, CNRS (UMR 8551), Université Pierre et Marie Curie, Université D. Diderot, 75231 Paris Cedex 05, France

**Table of contents**

## A Phase structure of the molecular eigenstates disregarding the spin degree of freedom

The polariton benzene molecule employed in our studies is made out of six overlapping cylindrical micropillars. In a single micropillar the three-dimensional confinement of polaritons results in discrete energy levels whose spacing is determined by the diameter of the molecule and the exciton-photon detuning. In our structure, with micropillars of 3 µm and zero photon-exciton detuning, the splitting[1] between the lowest and first excited state is on the order of 2 meV. The spatial overlapping of the micropillars creates a coupling between the ground states of adjacent pillars of 0.3 meV, resulting in the molecular energy levels $\epsilon_n$, $n = 0,1,2,3$, accessible in energy-resolved photoluminescence at low power as shown in Fig. 1b. The value of 0.3 meV was measured in two coupled micropillars with the same diameter and overlap as the micropillars in the hexagonal molecule [2]. In the actual hexagonal molecule, next-nearest neighbour tunnelling also plays a role in the energy difference between the $\epsilon_n$ energy levels, resulting in a splitting smaller than 0.3 meV between $\epsilon_0$ and $\epsilon_1$. This smaller splitting is well reproduced by the 2D Schrödinger equation simulations shown in Supplementary Figure 6.

Spin orbit coupling effects result in fine structure splittings that are smaller than the emission linewidth observed in the spontaneous emission regime. These effects can only be evidenced in the condensation regime under high excitation density. In this section of the Supplementary Information we concentrate in the low power emission and we disregard the spin.

The form of the molecular wavefunctions can be calculated with a tight binding approach in which the system is described by six sites with nearest-neighbour tunnelling. In this simple approximation, the stationary wavefunction of the $n^{th}$ level takes the form:

$$|\psi_n^l\rangle = \sum_k 6^{-1/2} e^{i\phi_{l,k}} |k\rangle$$

with

$$\phi_{l,k} = l\pi k/3,$$

where $|k\rangle$ is a wavefunction spatially localised in the $k = 0,\ldots,5$ micropillar. All the eigenfunctions present equal probability amplitudes in the six sites, resulting in equal emission intensities at the centre of each micropillar, as observed in Fig. 1d. The main difference is in the pillar to pillar relative phase, which can be parameterised by the angular momentum quantum number $l = \pm n$. The ground state presents a constant phase over all the pillars ($\phi_{0,k} = 0$), the levels $n = 1$ and $n = 2$ are both doubly degenerate with eigenvalues $l = +1, -1$ and $l = +2, -2$, respectively. In these eigenstates the phase of the wavefunction changes from 0 to $\pm 2\pi l$ when circumventing the molecule, defining a phase vortex of charge $\pm l$. The $n = 3$ state is non-degenerate (the periodicity of the effective potential creates a combination of the states $l = \pm 3$, with a phase that changes by $\pi$ from pillar to pillar).

The phase structure we just described can be accessed experimentally by an interferometric experiment at low excitation density (conditions of Fig. 1). We use the technique explained in the Methods: the emission from the ensemble of the molecule interferes with a magnified image of the emission from one of the micropillars, which provides a flat phase reference. Both images reach the entrance slit of an imaging spectrometer at an angle. In the spectrometer, each emission line is resolved in energy reaching a different position in the CCD camera. Thus, in the CCD, we observe the interference between the reference phase and the emission from the whole molecule, for each energy level. By performing an energy-resolved, two-dimensional, real-space tomography we can reconstruct the real-space interferometric images corresponding to the emission from each of the $\epsilon_0, \ldots, \epsilon_3$ energy levels. These images are shown in Supplementary Figure 1a-d in the conditions of Fig. 1. We then perform a Fourier Transform and filter out the off diagonal elements of the real and imaginary part. We Fourier transform back the filtered images and we obtain both the visibility of the fringes and the spatial changes of the phase, shown in Supplementary Figure 1e-l. More details of this technique can be found in Ref. [3].

The ground state, $\epsilon_0$, non-degenerate, shows a homogeneous relative phase at the centre of the pillars, as can be seen by following the dotted line in Supplementary Figure 1l (the radial variation of the phase is an artefact arising from the geometry of the interferometric experiment).

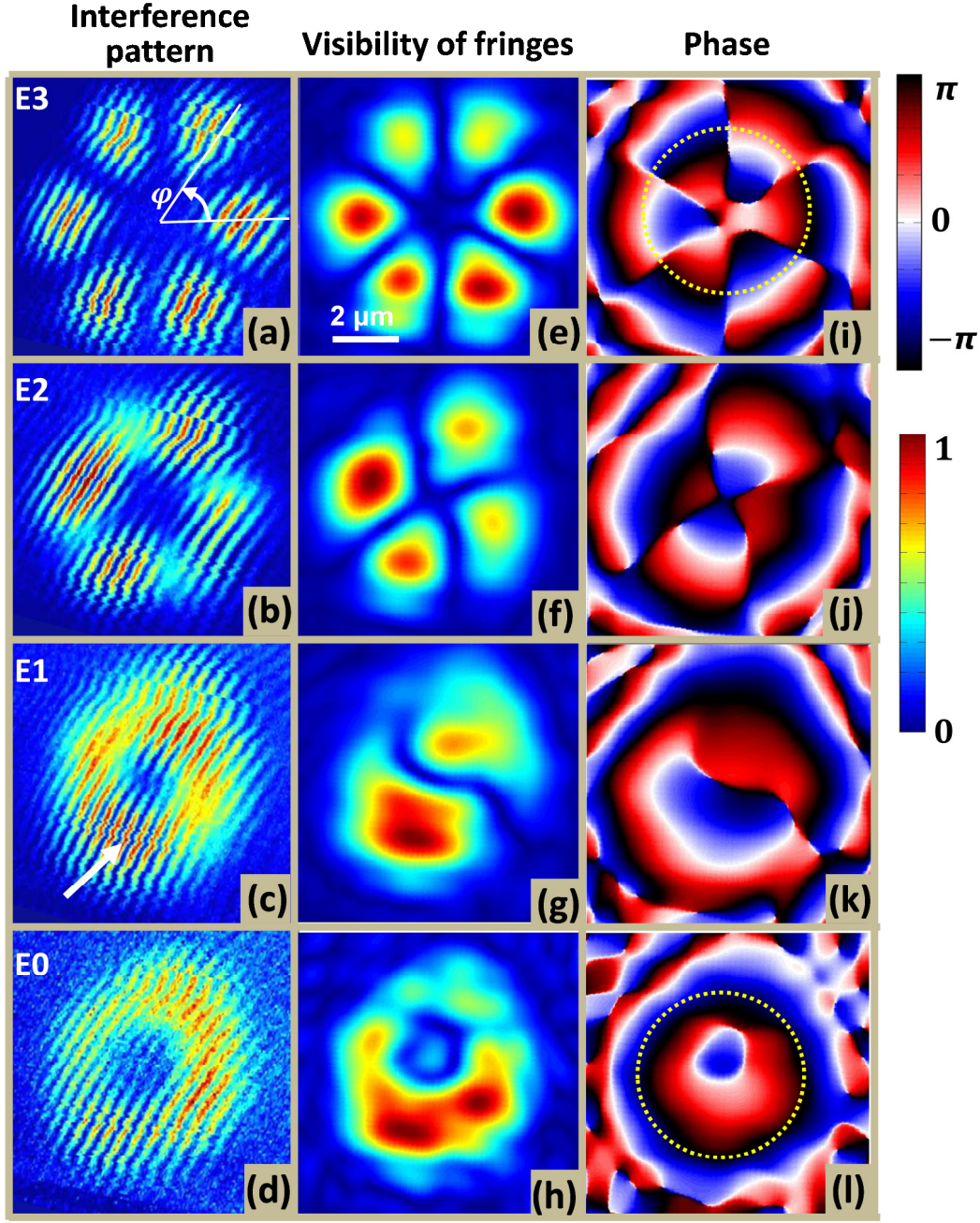

**Supplementary Figure 1.** Interference (a-d), visibility of fringes (e-h) and extracted phase (j-l) of the $n = 0,1,2,3$ levels measured in the spontaneous emission regime at low excitation density (conditions of Fig. 1 in the main text).

The $\epsilon_1$ energy level is double degenerate, with the emission being a superposition of states with $l = +1, -1$: $|\psi_1\rangle = \sum_k 12^{-1/2}\, e^{+i\pi k/3}|k\rangle + 12^{-1/2}\, e^{-i\pi k/3 + i\theta}|k\rangle$, where $\theta$ is a relative phase whose value, in the spontaneous emission regime, changes randomly in time. The interference pattern shown in Supplementary Figure 1c arises from the superposition between the emission from the whole molecule and a reference corresponding to an enlarged image of the micropillar located at the arrow ($k = 0$). In our time integrated images, interference fringes are only visible at the location of the reference pillar ($k = 0$, self-interference) and the one located in front ($k = 3$), which has a fixed relative phase difference $\pi$ for both $l = +1, -1$ substates. This phase difference is well observed in Supplementary Figure 1k. At intermediate positions, the phase difference respect to the emission from the $k = 0$ micropillar is scrambled when integrating the images due to the random values of $\theta$, resulting in a low visibility (Supplementary Figure 1g).

A similar situation is observed for the degenerate states $l = +2, -2$ ( $\epsilon_2$ ): $|\psi_2\rangle = \sum_k 12^{-1/2} e^{+i2\pi k/3}|k\rangle + 12^{-1/2} e^{-i2\pi k/3+i\theta}|k\rangle$. As the phase now increases with the angle twice as fast, there are four positions in the molecule in which the phase difference with respect to the reference $k = 0$ micropillar is the same for both substates (an integer multiple of $\pi$), thus resulting in a high fringe visibility (Supplementary Figure 1b,f). For those four positions, a phase jump of $\pi$ is well observed (Supplementary Figure 1j).

The non-degenerate $n = 3$ state shows a high fringe visibility for all the micropillars (Supplementary Figure 1a,e), with a phase jump of $\pi$ from pillar to pillar (follow dotted lines in Supplementary Figure 1i).

**B Measurement of the polarisation structure of the $n = 1$ and $n = 2$ condensed states**

The polarisation of light emitted at a particular point of the sample can be described in the paraxial approximation as in Born and Wolf[4]. We can define an arbitrary polarisation state in the following way:

$$E_x = a_1(\boldsymbol{r})cos(\omega\tau + \delta_1(\boldsymbol{r}))$$

$$E_y = a_2(\boldsymbol{r})cos(\omega\tau + \delta_2(\boldsymbol{r}))$$

where $a_{1,2}(\boldsymbol{r})$ describes the amplitude of the electric field along the $x$ and $y$ directions in the plane of the molecule at a given point $\boldsymbol{r}$, $\omega$ is the frequency of light and $\delta_{1,2}(\boldsymbol{r})$ are fixed phases for each component. The polarisation state is fully determined by the ratio $a_1(\boldsymbol{r})/a_2(\boldsymbol{r})$ and by the phase difference $\delta_2(\boldsymbol{r}) - \delta_1(\boldsymbol{r})$.

The condition for linear polarisation is $\delta = \delta_2(\boldsymbol{r}) - \delta_1(\boldsymbol{r}) = m\pi$, with $m = 0, \pm 1, \pm 2, ...$, with the ratio $a_1(\boldsymbol{r})/a_2(\boldsymbol{r})$ defining the direction of polarisation. Circular polarisation is obtained under the condition $a_1(\boldsymbol{r}) = a_2(\boldsymbol{r})$, $\delta = \delta_2(\boldsymbol{r}) - \delta_1(\boldsymbol{r}) = m\pi/2$, with $m = \pm 1, \pm 3, \pm 5, ...$ Other values of $a_1(\boldsymbol{r})/a_2(\boldsymbol{r})$ and $\delta$ result in elliptical polarisations.

A basis of interest is that of circular polarisation:

$$\boldsymbol{E}_{\sigma+} = a_{\sigma+}(\boldsymbol{r})cos(\omega\tau + \delta_{\sigma+}(\boldsymbol{r}))\boldsymbol{x} - i\, a_{\sigma+}(\boldsymbol{r})cos(\omega\tau + \delta_{\sigma+}(\boldsymbol{r}))\boldsymbol{y}$$

$$\boldsymbol{E}_{\sigma-} = a_{\sigma-}(\boldsymbol{r})cos(\omega\tau + \delta_{\sigma-}(\boldsymbol{r}))\boldsymbol{x} + i\, a_{\sigma-}(\boldsymbol{r})cos(\omega\tau + \delta_{\sigma-}(\boldsymbol{r}))\boldsymbol{y}$$

In this basis, a linear polarised state corresponds to $a_{\sigma+}(r) = a_{\sigma-}(r)$; the phase difference $\delta_\sigma(r) = \delta_{\sigma+}(r) - \delta_{\sigma-}(r)$ sets the direction of the linear polarisation: $\theta = -(\delta_{\sigma+}(r) - \delta_{\sigma-}(r))/2$, with $\theta$ the clockwise angle from the $x$ direction.

Experimentally, the polarisation state of the emission can be fully described by the Stokes coefficients:

$$S_0 = I_{tot}$$

$$S_1 = \frac{I_x - I_y}{S_0}$$

$$S_2 = \frac{I_{+45} - I_{-45}}{S_0}$$

$$S_3 = \frac{I_{\sigma+} - I_{\sigma-}}{S_0}$$

where $I_{x,y,+45,-45,\sigma+,\sigma-}$ is the emitted intensities when detecting the linearly polarised emission along the x, y axis, the +45º, -45º directions with respect to the x axis, and the $\sigma+, \sigma-$ circularly polarisation, respectively. $S_1, S_2, S_3$, are defined such that they correspond to the degree of polarisation along the same axes. The relation to the $a_{\sigma+}/a_{\sigma-}$ ratio and $\theta$ is:

$$\frac{a_{\sigma+}}{a_{\sigma-}} = \frac{S_3 + 1}{S_3 - 1}$$

$$\theta = \frac{1}{2}\arctan\left(\frac{S_2}{S_1}\right)$$

Graphically, the polarisation state corresponds to a point in the Poincaré sphere determined by the $S_1, S_2, S_3$ axis:

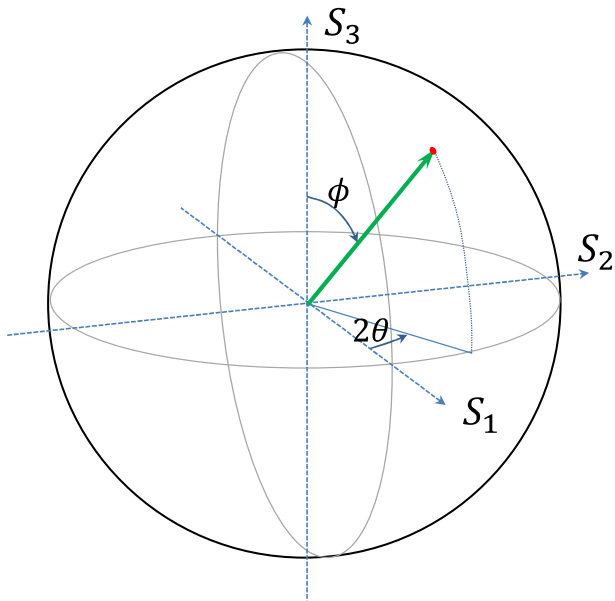

**Supplementary Figure 2**. Poincaré sphere representing the polarisation state of the emitted light. Note that $\phi = \arccos S_3$.

In order to reconstruct the linear polarisation patterns shown in Figs. 3g and 4g, we measure the Stokes coefficients $S_1$ and $S_2$ by using a linear polariser in combination with a half-waveplate. Due to constraints in the experimental set-up, in order to reconstruct the linear polarisation map shown in Figs. 3g and 4g we measure $S_1$ and $S_2$ along the polarisation axes oriented 22.5º/112.5º and 67.5º/157.5º with respect to the x direction (horizontal) in Figs. 3 and 4. We can redefine the coefficients $S_1$ and $S_2$ along those axes:

$$S_1' = \frac{I_{22.5} - I_{112.5}}{S_0}$$

$$S_2' = \frac{I_{67.5} - I_{-157.5}}{S_0}$$

$S_3$ is measured by using a quarter-waveplate and a linear polariser.

Supplementary Figure 3 and 4 show the polarisation emission filtered in the different projections needed to reconstruct $S_1'$, $S_2'$, and $S_3$ corresponding to the situations described in Figs. 3 and 4 of the main text, respectively. The angle $\alpha$ shows the orientation of the polarizers used to analyse the emission with respect to the horizontal axis of the molecule, as defined in the inset. From images (a) through (d) in Supplementary Figures 3 and 4 we can extract the direction of the linear polarisation of the emission plotted in Figs. 3g and 4g. The angle $\theta$ setting the direction of linear polarisation at each point in the plane of the figures is calculated from the following expression: $\tan[2 \cdot (\theta + 22.5°)] = S_2'/S_1'$ where $\theta$ increases counterclockwise and 0 corresponds to the horizontal positive direction. Note that the shallow diagonal traces observed in Supplementary Figures 3 and 4 arise from an artefact due to the presence of a neutral density filter in the detection path.

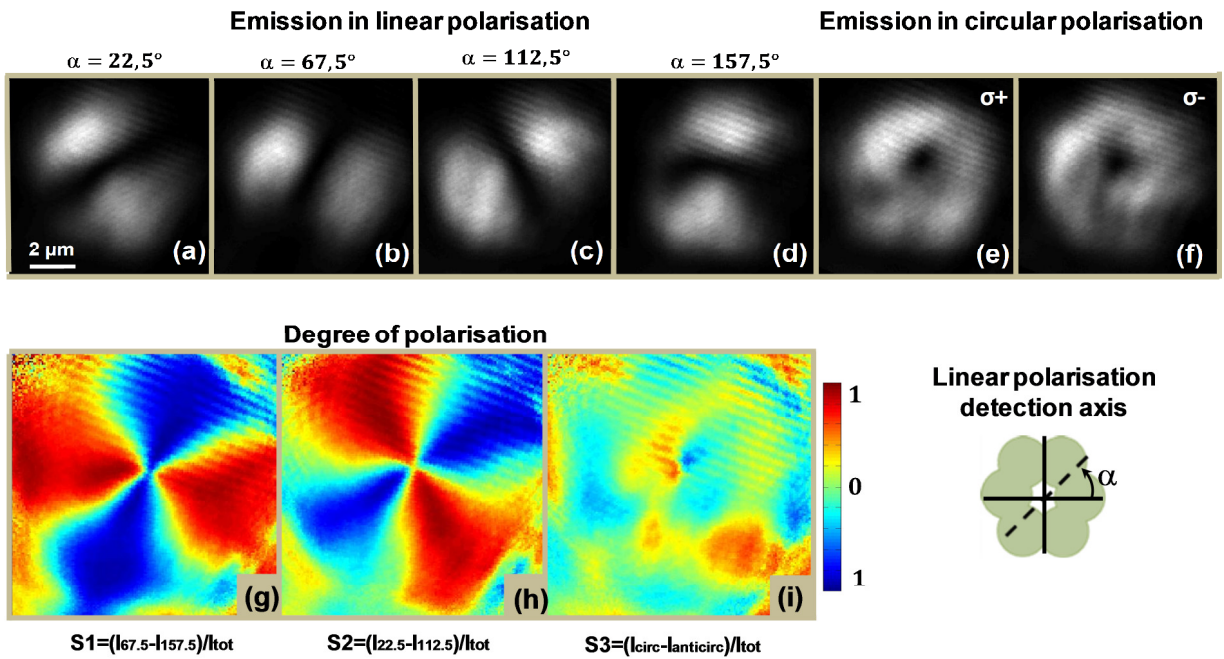

**Supplementary Figure 3.** Linear polarisation tomography of the $n = 1$ condensed state shown in Fig. 3 of the main text. From the degree of linear polarisation reported in (g)-(h), we extract the overall linear polarisation direction shown in Fig. 3g.

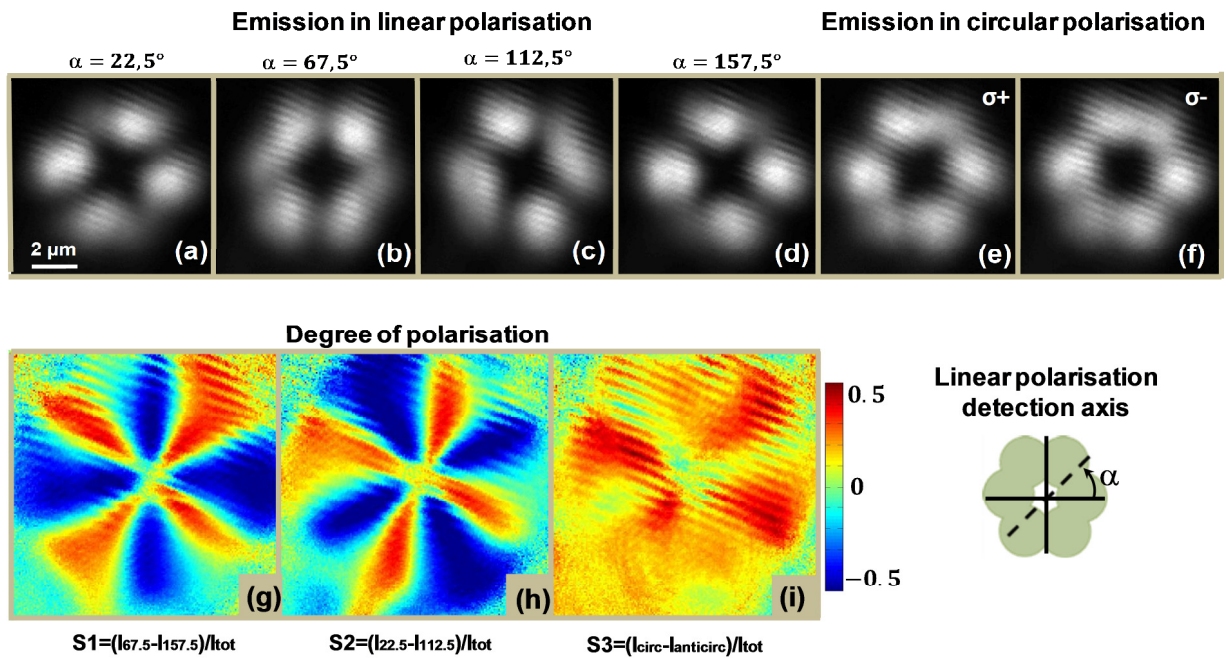

**Supplementary Figure 4.** Linear polarisation tomography of the $n = 1$ condensed state shown in Fig. 4 of the main text. From the degree of linear polarisation reported in (g)-(h), we extract the overall linear polarisation direction shown in Fig. 4.

## C Condensation kinetics

In our structure, we observe two condensation thresholds at two different excitation powers as shown in Fig. 3h. This can be more clearly observed when looking at the angle resolved spectra at different powers, shown in Supplementary Figure 5.

Between 7 and 17mW, below the first threshold, we observe a blueshift related to the continuous increase of reservoir excitons in the system. At 25 mW, condensation takes place in the E2 level. At higer power, 57mW, we observe simultaneous condensation in the E1 and E2 levels. Finally, above 84mW only the E1 level remains highly occupied. Note that the redshift observed for the E1 level between 57 and 110mW arises from the heating of the sample due to the large absorbed optical density.

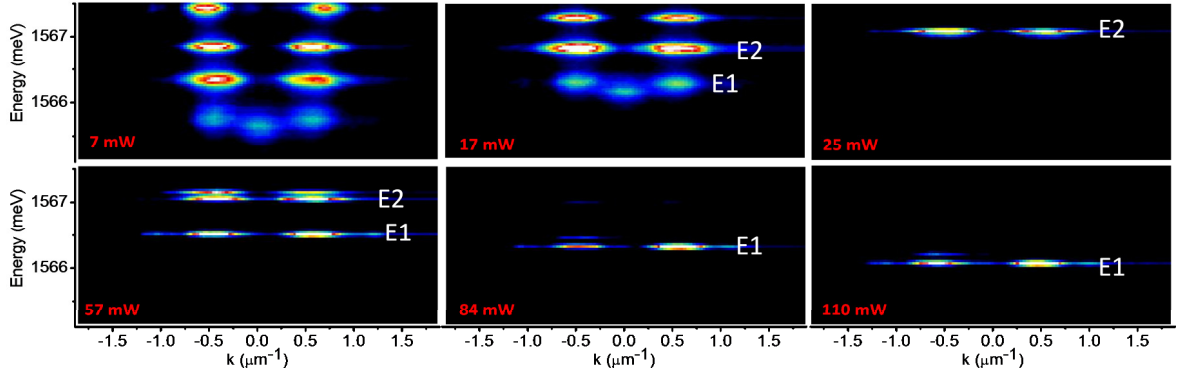

**Supplementary Figure 5**. Experimental spectra at different excitation densities.

In an open dissipative system with pumping and lifetime the Bose Einstein condensation does not necessarily occur in the ground state since relaxation kinetics has to be taken into account[5]. Within the simplest approximation, the equation describing the occupation of a confined state reads: given by

$$\frac{dN}{dt} = W_{in}(N+1) - N(W_{out} + 1/\tau), \qquad (1)$$

where $\tau$ is the state lifetime. $W_{in,out}$ are the scattering rates toward and outward the state. They verify:

$$W_{in} = W_{out} e^{-E/k_b T} = W Y_{cR} e^{-E/k_b T},$$

where $Y_{cR}$ is the overlap integral between the state density and the exciton reservoir distribution. $W$ depends on the system parameters, but since the main relaxation mechanism is based on exciton-exciton interaction, $W$ is a growing function of the carrier density and of the pumping power. Equation 1 can be recast as

$$\frac{dN}{dt} = W_{in}(N+1) - N(\tilde{W}_{out})$$

where

$W_{in} = \tilde{W}_{out} e^{-E'/k_b T} = W Y_{cR} e^{-E/k_b T}$ which yields

$$E' = E + k_b T \ln\left(1 + \frac{1}{WY_{cR}\tau}\right) \qquad (2)$$

$W$, $Y$ and $\tau$ are in general functions of $E$. The meaning of the previous development is that the state occupation in our pump-dissipative system can be expressed as thermal-like distribution function if one defines a new energy scale that includes the effect of particle life time. This is what is expressed in Eq. 2, where we can see that the state with the lower effective energy $E'$ is not necessarily the original ground state of the system. For instance, an excited state with a long lifetime $\tau$ can become the state with lowest effective $E'$, and thus, the most favoured for condensation to take place. The exact value of $E'$ depends on the relaxation efficiency, pumping power, overlap integral between the reservoir and the state, lifetime. In the rest of the text we are going to analyse which is the most favoured state in the molecule depending on the excitation conditions. The increase of $W$ with pumping reduces the impact of the kinetics on the determination of the ground state.

**_Lifetime_** As a first approximation, the polariton lifetime in this system depends on the absolute value of the angular momentum. Indeed, the decay of the particles is given by the extension of the wave function out of the structure. This extension is maximal for bound states $l = 0$, it decreases for higher values of $l$ and is minimal for the anti-bound states[6] $l = 3$ with a variation which we estimated numerically of the order of 20 %. As a result, condensation in states with $l = 0$ is not favoured by kinetics because of their short lifetime.

There are two additional mechanisms that result in the increase of the polariton lifetime with increasing energy. First of all the exciton content increases with energy, resulting in an enhanced lifetime. Second, the wavefunction of higher energy modes presents zeroes at the constrictions between the pillars, where the defect density is higher. Thus, higher energy modes are more protected against the non-radiative losses associated to defects. The combination of these effects results in condensation in the E3 energy level with the lowest threshold.

**_Overlap with the reservoir, spin-anisotropic interaction_**. As said above, the degenerate state can be represented either, as spatially homogeneous circularly polarized states or as linearly polarized spatially inhomogeneous states. The spin-anisotropic interaction of polaritons makes that it is favorable energetically for a condensate to be linearly polarized[7]. On the other hand, linear combination of states resulting in amplitude inhomogeneities show a reduced overlap with the excitonic reservoir, which is homogeneous all over the molecule in our excitation conditions. For these reasons, the states which favour condensation are the spatially homogeneous states resulting from the coupling between $l_+ = -1$ and $l_- = 1$ and $l_+ = 2$, $l_- = -2$.

***Summary.*** At low pumping, condensation is expected to occur on the more kinetically favoured states, namely the states resulting from the coupling $l_+ = 2$, $l_- = -2$. It is also reasonable to expect that the system will choose the lowest of these two states. Going to higher power, the effective energies of the state evolve and the effective ground state should become the lowest of the two states resulting from the coupling between $l_+ = -1$, $l_- = 1$, namely, the state with an azimuthal polarisation.

***Numerical simulations.*** In order to confirm this finding, we use the model based on self-consistent coupled semi-classical Boltzmann and nonlinear Schrödinger equations[2], and find a good agreement with the experimental measurements.

To be able to carry out a direct comparison with the experiment, we calculate the emission from the quantized states of our benzene molecule in the reciprocal space, taking into account the occupation numbers found from the simulations described above. The results are shown in Supplementary Figure 6. Below threshold (left panel), all states are approximately equally populated. Main emission comes from the ground state which can be identified by a maximum of emission at k=0. At higher pumping (right panel), condensation occurs at the most favoured state which is the state with $l = 1$ and azimuthal polarization. The emission from this state dominates the spectrum.

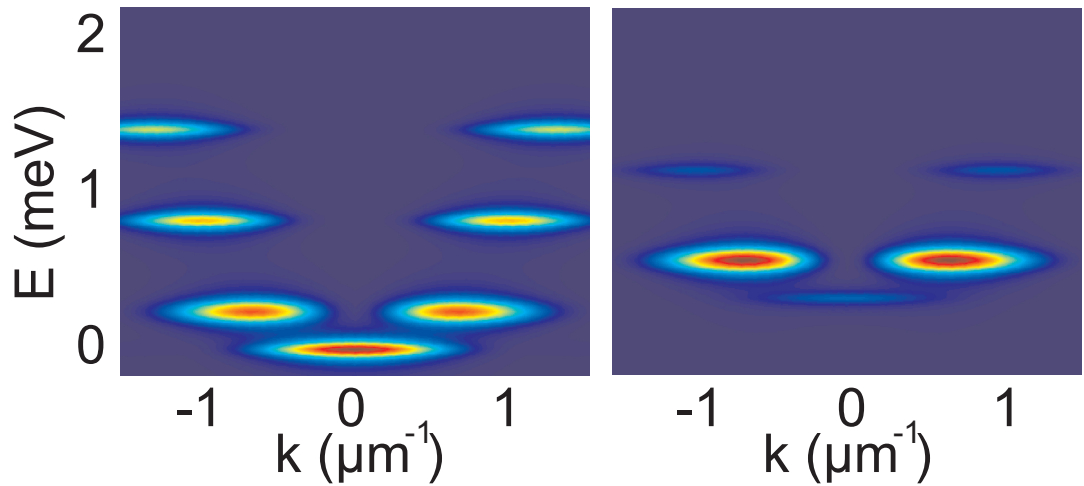

**Supplementary Figure 6**. Simulated emission pattern in the reciprocal space below (left panel) and above threshold (right panel). Condensation occurs on the state with angular momentum 1 and azimuthal polarization.

# Supplementary Information for
# "Engineering spin-orbit coupling for photons and polaritons in microstructures"
# D.- Model

V.G. Sala,[1,2] D. D. Solnyshkov,[3] I. Carusotto,[4] T. Jacqmin,[1] A. Lemaître,[1] H. Terças,[3]
A. Nalitov,[3] M. Abbarchi,[1,5] E. Galopin,[1] I. Sagnes,[1] J. Bloch,[1] G. Malpuech,[3] and A. Amo[1]

[1]*Laboratoire de Photonique et Nanostructures, CNRS/LPN, Route de Nozay, 91460 Marcoussis, France*
[2]*Laboratoire Kastler Brossel, Université Pierre et Marie Curie, École Normale Supérieure et CNRS,
UPMC Case 74, 4 place Jussieu, 75252 Paris Cedex 05, France*
[3]*Institut Pascal, PHOTON-N2, Clermont Universit, Universit Blaise Pascal,
CNRS, 24 avenue des Landais, 63177 Aubire Cedex, France*
[4]*INO-CNR BEC Center and Dipartimento di Fisica, Università di Trento, I-38123, Povo, Italy*
[5]*Laboratoire Pierre Aigrain, École Normale Supérieure,
CNRS (UMR 8551), Université Pierre et Marie Curie,
Université D. Diderot, 75231 Paris Cedex 05, France*
(Dated: April 29, 2014)

## I. TIGHT-BINDING DERIVATION OF THE SPIN-ORBIT HAMILTONIAN OF PHOTONIC BENZENE

In this section we present the tight-binding derivation of the spin-orbit Hamiltonians (1) and (2) of the main text. We consider a chain of cylindrical micropillar cavities arranged at the vertices of a regular polygon of $M$ sides, as sketched in Fig. 1 for the $M = 6$ case of the benzene experiments. The polygon is assumed to sit on the $x - y$ plane.

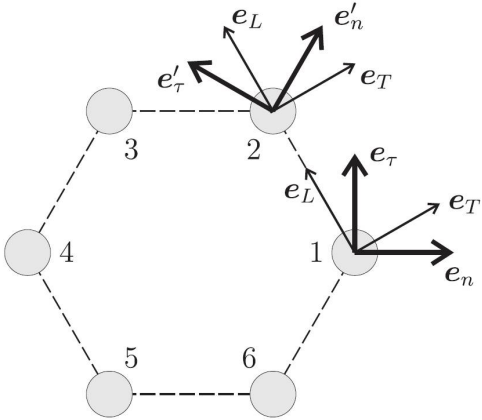

FIG. 1. Sketch of the system under consideration.

On each site, a single orbital mode is available, with an approximately cylindrically symmetric wavefunction. Each orbital mode has a twofold spin degeneracy corresponding to the two polarization directions on the $x$ and $y$ directions. On the $j$th site centered at position $\mathbf{R}_j$, the vector electric (or polaritonic) field operator has the form:

$$\hat{\vec{\mathbf{E}}}(\mathbf{r}) = \phi(\mathbf{r} - \mathbf{R}_j)[\mathbf{e}_x \hat{a}_{j,x} + \mathbf{e}_y \hat{a}_{j,y}] =$$
$$= \phi(\mathbf{r} - \mathbf{R}_j)[\mathbf{e}_{\sigma_+} \hat{a}_{j,\sigma_+} + \mathbf{e}_{\sigma_-} \hat{a}_{j,\sigma_y}], \quad (1)$$

where the $\hat{a}_x$, $\hat{a}_y$ and $\hat{a}_{\sigma_+}$, $\hat{a}_{\sigma_-}$ are two among the possible basis on which the vector electric field can be expanded. $\mathbf{e}_{x,y}$ is a cartesian basis, while $\mathbf{e}_{\sigma_\pm} = (\mathbf{e}_x \pm i\mathbf{e}_y)/\sqrt{2}$ is the circular basis.

In the following, it will be convenient to use the compact and basis-independent vector notation

$$\hat{\mathbf{a}}_i = \mathbf{e}_x \hat{a}_{j,x} + \mathbf{e}_y \hat{a}_{j,y} = \mathbf{e}_{\sigma_+} \hat{a}_{j,\sigma_+} + \mathbf{e}_{\sigma_-} \hat{a}_{j,\sigma_y}. \quad (2)$$

For both the cartesian and the circular basis, the commutators satisfy bosonic commutation rules.

For each link connecting the $j \to j+1$ sites, we can define the real unit vectors $\mathbf{e}_L^{(j)}$ and $\mathbf{e}_T^{(j)}$ respectively parallel and orthogonal to the link direction $\mathbf{R}_{i+1} - \mathbf{R}_i$. The main assumption of our model is that tunneling along the $j \to j+1$ link occurs with different amplitudes $t_L$ and $t_T$ for photons polarized along the $\mathbf{e}_L^{(i)}$ and $\mathbf{e}_T^{(i)}$ directions, respectively. According to S. Michaelis de Vasconcellos et al., Appl. Phys. Lett. **99**, 101103 (2011), $|t_L| \gtrsim |t_T|$.

In second quantization terms, the many-body Hamiltonian then reads:

$$H = -\sum_{j=1}^{M}\{[\hbar t_L (\hat{\mathbf{a}}_{j+1}^\dagger \cdot \mathbf{e}_L^{(j)})(\mathbf{e}_L^{(j)\dagger} \cdot \hat{\mathbf{a}}_j) + $$
$$+ \hbar t_T (\hat{\mathbf{a}}_{j+1}^\dagger \cdot \mathbf{e}_T^{(j)})(\mathbf{e}_T^{(j)\dagger} \cdot \hat{\mathbf{a}}_j)] + \text{h.c.}\}. \quad (3)$$

Physically, the $\mathbf{e}_{L,T}^{(j)\dagger} \cdot \hat{\mathbf{a}}_j$ expression selects the component of the vector $\hat{\mathbf{a}}_j$ field on the $j$th side along the $\mathbf{e}_{L,T}^{(j)}$ unit vector. Given the periodic boundary conditions of the system, in the sum the $(M+1)$th site has to be identified with the 1st and the 0th with the $M$th.

As the many-body Hamiltonian (3) does not involve any interaction terms, it is straightforward to derive a Schrödinger equation for the single-particle ℂ-number vector wavefunction $\boldsymbol{\alpha}_j$ that defines states on the one-body subspace:

$$|\psi_1\rangle = \sum_j [\boldsymbol{\alpha}_j \cdot \hat{\mathbf{a}}_j^\dagger] |\text{vac}\rangle \quad (4)$$

As usual, the Schrödinger equation can be obtained from the Heisenberg equation for the field operators $\hat{a}_j$

$$i\hbar \frac{d}{dt}\hat{a}_j = [\hat{a}_j, H] \quad (5)$$

by replacing the field operators $\hat{a}_j$ with the one-body wavefunction $\alpha\alpha_j$. In this way, one obtains

$$i\frac{d\boldsymbol{\alpha}_j}{dt} = t_L \mathbf{e}_L^{(j-1)}(\mathbf{e}_L^{(j-1)\dagger}\cdot\boldsymbol{\alpha}_{j-1}) + t_T \mathbf{e}_T^{(j-1)}(\mathbf{e}_T^{(j-1)\dagger}\cdot\boldsymbol{\alpha}_{i-1}) +$$
$$+ t_L \mathbf{e}_L^{(j)}(\mathbf{e}_L^{(j)\dagger}\cdot\boldsymbol{\alpha}_{j+1}) + t_T \mathbf{e}_T^{(j)}(\mathbf{e}_T^{(j)\dagger}\cdot\boldsymbol{\alpha}_{j+1}). \quad (6)$$

If $M$ is even, the symmetry operation $\mathcal{S}$ defined as:

$$\mathcal{S}^\dagger \hat{a}_j \mathcal{S} = (-1)^j \hat{a}_j \quad (7)$$

anticommutes with the Hamiltonian

$$\mathcal{S}^\dagger H \mathcal{S} = -H. \quad (8)$$

As a result, if $|\psi\rangle$ is an eigenstate of the Hamiltonian $H$ of energy $E$, then $\mathcal{S}|\psi\rangle$ is an eigenstate of opposite energy $-E$. At the single-particle level, the action of the operator $S$ is to invert the sign of the field on the odd sites

$$(\mathcal{S}\boldsymbol{\alpha})_j = (-1)^j \boldsymbol{\alpha}_j, \quad (9)$$

so that the tunneling energy of each link changes sign.

The standard translation operator $T$ is defined as

$$T^\dagger \hat{a}_j T = \hat{a}_{j+1} : \quad (10)$$

physically, the photon field on the translated state $T|\psi\rangle$ at site $j$ is equal to the original field on the site $j+1$. In the general case $t_L \neq t_T$, the operator $T$ does not commute with the Hamiltonian: translation changes the relative orientation of the photon field $\hat{a}_j$ and the link frame defined by the pair $\mathbf{e}_{L,T}^{(j)}$.

One has to define a new translation operator $\tilde{T}$ as follows

$$T^\dagger \hat{a}_j T = \mathcal{R}_{-2\pi/M} \hat{a}_{j+1} : \quad (11)$$

translation by one site has to be combined with a rotation by an angle $-2\pi/M$ so to compensate the different orientation of the link. This new operator $\tilde{T}$ commutes with the Hamiltonian $H$. It is immediate to check that after a full round trip around the chain, it gives back the identity $\tilde{T}^M = \mathbf{1}$. In physical terms, the operator $\tilde{T}$ describes discrete rotations by $2\pi/M$ around the chain, and therefore corresponds to the total angular momentum of the state.

Let's now restrict to the single-particle space. As $\tilde{T}$ commutes with $H$, eigenstates can be found with well defined quantum number $k$ as compared to generalized translations, that is, such that

$$\tilde{T}|\psi\rangle = e^{2\pi i k/M}|\psi\rangle \quad (12)$$

with $k = 1, \ldots, M$. It is important to note that the wavevector $k$ does not straightforwardly correspond to the orbital part of the angular momentum. This is easy to see on $\sigma_\pm$ circularly polarized states such that

$$\boldsymbol{\alpha}_j = \alpha_j \begin{pmatrix} 1 \\ \pm i \end{pmatrix}. \quad (13)$$

In this case, if the wavefunction has orbital angular momentum is $l \cdot \pi/3$ (again $l = 1, \ldots, M$), $\alpha_j = \alpha_o e^{ijl\pi/3}$, the total angular momentum of the state

$$\boldsymbol{\alpha}_j = \alpha_o e^{ijl\pi/3}\begin{pmatrix} 1 \\ \pm i \end{pmatrix}. \quad (14)$$

contains a spin contribution, $k = l \pm 1$. On the other hand, an experiment where $\sigma_\pm$ light is selected, is sensitive to the orbital wavefunction only.

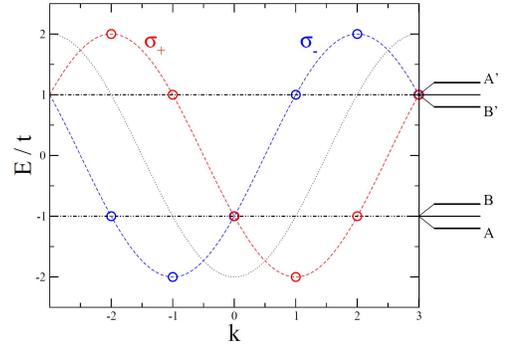

FIG. 2. Sketch of the eigenstates of a $M = 6$ photonic benzene system. Red (blue) points indicate the $\sigma_+$ ($\sigma_-$) polarized states. States are labeled in terms of the total angular momentum $k$.

Let's begin to diagonalize $H$ in the $t_L = t_T = t$ case. In this case, on each site one has

$$\mathbf{e}_L^{(j)} \times \mathbf{e}_L^{(j)\dagger} + \mathbf{e}_T^{(j)} \times \mathbf{e}_T^{(j)\dagger} = \mathbf{1}. \quad (15)$$

The energy of the state is therefore given by the orbital energy only, equal to

$$E_l = -2\hbar t \cos(2\pi l/M). \quad (16)$$

Even though the spin decouples from the orbital motion, it is instructive to see the behavior of the generalized translation operators $\tilde{T}$. The circularly polarized states are eigenstates of the rotation operator

$$\mathcal{R}_\theta \sigma_\pm = e^{-i\theta}\sigma_\pm. \quad (17)$$

As a result, the energy of $\sigma = \sigma_\pm$ polarized states of total angular momentum $k$ is

$$E(k, \sigma) = -2\hbar t \cos[2\pi(k - \sigma)/M] \quad (18)$$

and its dispersion is sketched in Fig. 2. The presence of a spin-orbit coupling term is apparent in (18), where the dispersion of the $\sigma_\pm$ is laterally shifted by $\Delta k = \pm 1$. In Hamiltonian terms, we can write

$$H = \sum_{k,\sigma} E(k,\sigma)\, \hat{b}^\dagger_{k,\sigma} \hat{b}_{k,\sigma}, \qquad (19)$$

where

$$\hat{b}_{k,\sigma} = \frac{1}{\sqrt{M}} \sum_j e^{-2\pi i(k-\sigma)j/M} \hat{a}_{j,\sigma} \qquad (20)$$

is the destruction operator for a photon of total angular momentum $k$ and spin $\sigma$

In the general case $t_L \neq t_T$, the orbital angular momentum $l$ is no longer a good quantum number, but $k$ remains so: the $t_L - t_T$ coupling does not break rotational invariance. To qualitatively understand the degeneracies of the eigenstates in the experimentally relevant case $|t_L - t_T| \ll t_{L,T}$ it is useful to draw the dispersion of the $\sigma_\pm$ states as a function of $k$ and include the effect of $t_L - t_T$ only at the crossing points of the bands. As one can see in Fig. 2, for even values of $M$ such crossings only occur at $k = 0, M/2$. For all other values, the bands are non-degenerate.

As a result, the ground state is two-fold degenerate and is spanned by the eigenstates $|\sigma_+, k=1\rangle$ and $|\sigma_-, k=-1\rangle$ (as usual, because of the discrete rotational symmetry the quantum number $k$ is only defined modulo $M$, so $k=-1$ is equivalent to $k=M-1$). Its energy is $\simeq -2t$. Correspondingly, thanks to the $\mathcal{S}$ symmetry mentioned above (the $\mathcal{S}$ operation sends a state with orbital momentum $l$ into a state of orbital momentum $l + M/2$), the highest energy state is two-fold degenerate at energy $\simeq 2t$ and is spanned by $|\sigma_+, k=M/2+1\rangle$ and $|\sigma_-, k=-M/2-1\rangle$.

The first excited manifold at $E \simeq -2\hbar t \cos(2\pi/M) = -\hbar t$ was four-fold degenerate for $t_L = t_T$. The two external states $(\sigma_+, k=2)$ and $(\sigma_-, k=-2)$ have no other available state at the same $k$, so remain degenerate at $E = -\hbar t$. On the otehr hand, the two other states $|\sigma_+, k=0\rangle$ and $|\sigma_-, k=0\rangle$ can be mixed by the angular momentum conserving $t_L - t_T$ terms.

To understand the form of the new eigenstates, it is useful to give an explicit expression for the $\Delta t = t_L - t_T$ terms in the $\sigma_\pm$ basis; from now on we will assume for clarity $\Delta t > 0$. Rewriting the Hamiltonian

$$H = -\sum_{j=1}^M \Big\{ \frac{\hbar t}{2}[\hat{\mathbf{a}}^\dagger_{j+1} \hat{\mathbf{a}}_j + \hat{\mathbf{a}}^\dagger_j \hat{\mathbf{a}}_{j+1}] + \\ - \frac{\hbar \Delta t}{2}[(\hat{\mathbf{a}}^\dagger_{j+1} \cdot \mathbf{e}_L^{(j)})(\mathbf{e}_L^{(j)\dagger} \cdot \hat{\mathbf{a}}_j) + \\ - (\hat{\mathbf{a}}^\dagger_{j+1} \cdot \mathbf{e}_T^{(j)})(\mathbf{e}_T^{(j)\dagger} \cdot \hat{\mathbf{a}}_j)] + \text{h.c.}\Big\}. \quad (21)$$

and expressing it in terms of the $k$-space operators, one gets to

$$H = -\sum_{j=1}^M \Big\{ \frac{\hbar t}{2}[\hat{\mathbf{a}}^\dagger_{j+1} \hat{\mathbf{a}}_j + \hat{\mathbf{a}}^\dagger_j \hat{\mathbf{a}}_{j+1}] + \\ - \frac{\hbar \Delta t}{2} \sum_k \cos\Big(\frac{\pi k}{3}\Big) [\hat{a}^\dagger_{k,\sigma_-} \hat{a}_{k,\sigma_+} e^{-4\pi i/M} + \text{h.c.}]. \quad (22)$$

From this expression it is immediate to see that the $\Delta t$ term still conserves angular momentum and that the $|\sigma_+, k=0\rangle$ and $|\sigma_-, k=0\rangle$ give rise to new eigenstates at energies $E \simeq -2\hbar t \cos(2\pi/M) \pm \hbar \Delta t$. Given the phase factor in the last term of (28), the lowest eigenstate at energy $\hbar(-2t\cos(2\pi/M) - \Delta t)$ is of the form

$$\psi^{(A)} = \frac{1}{\sqrt{2}}[e^{2\pi i/M}|0,\sigma_+\rangle + e^{-2\pi i/M}|0,\sigma_-\rangle], \qquad (23)$$

while the highest at energy $\hbar(-2t\cos(2\pi/M) + \Delta t)$ has the form

$$\psi^{(B)} = \frac{1}{\sqrt{2}}[e^{2\pi i/M}|0,\sigma_+\rangle - e^{-2\pi i/M}|0,\sigma_-\rangle]. \qquad (24)$$

These are the upper and lower states sketched in Fig. 1f of the main text. It is interesting to get insight on the spatial polarization structure of the $\psi_{A,B}$ states. Replacing the explicit expression of the $|0,\sigma_\pm\rangle$ states, one obtains for the lower $A$ state

$$\psi_j^{(A)} = \frac{1}{2}[e^{-2\pi i(j-1)/M}\begin{pmatrix} 1 \\ i \end{pmatrix} + e^{2\pi i(j-1)/M}\begin{pmatrix} 1 \\ -i \end{pmatrix}] = \\ = \begin{pmatrix} \cos(2\pi(j-1)/M) \\ \sin(2\pi(j-1)/M) \end{pmatrix} \quad (25)$$

which is an azymuthal polarization: For instance, for $j=1$ the polarization is horizontal. Analogously for the higher $B$ state,

$$\psi_j^{(B)} = \frac{1}{2}[e^{-2\pi i(j-1)/M}\begin{pmatrix} 1 \\ i \end{pmatrix} - e^{2\pi i(j-1)/M}\begin{pmatrix} 1 \\ -i \end{pmatrix}] = \\ = i\begin{pmatrix} -\sin(2\pi(j-1)/M) \\ \cos(2\pi(j-1)/M) \end{pmatrix} \quad (26)$$

which is a radial polarization. These features are visible in the polarization patterns of the different eigenstates that are plotted in Fig. 3. In particular, the $A$ state is the one at $E = -1.002$ (we have used $\hbar t = 1$ and $\hbar \Delta t = 0.002$) on the third column (from left) of the top row. The $B$ state is the one at $E = -0.998$ on the second column of the second row.

The structure of the second highest energy manifold at $E = \hbar t$ (corresponding to Fig. 1g of the main text) is directly obtained via the $\mathcal{S}$ symmetry that sends the total angular momentum $k \to k + \pi/M$ while keeping the polarization pattern intact. As a result, the highest state $A'$ of this manifold will have a azimuthal polarization, while the lowest one $B'$ will have a radial polarization.

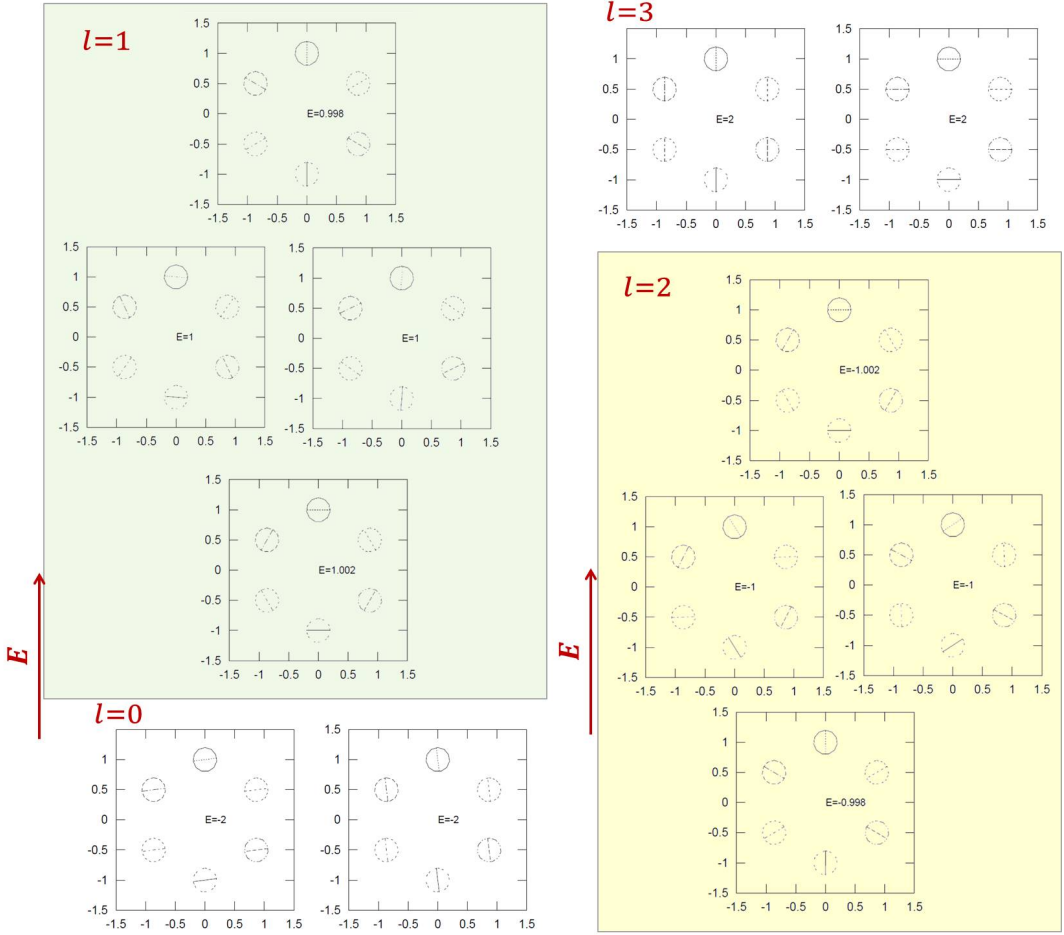

FIG. 3. Calculated polarization pattern for the different eigenstates of a photonic benzene molecule. Eigenstates are sorted for growing eigenenergy, indicated in the center of each panel (bottom-up on each column). The calculation is performed by diagonalizing the Hamiltonian (3) within the one-particle subspace. Parameters: $t_L = 1.001$, $t_T = 0.999$, $\Delta E = 0$.

This is again visible in Fig. 3 by comparing the state at $E = 1.002$ on the second column of the lowest row and the state at $E = 0.998$ on the third column of the second row.

In addition to the polarisation dependent tunnelling we have considered up to now $(t_L - t_T)$ our photonic structures may show an onsite splitting $\Delta E$ between modes linearly polarised azimuthal and radially with respect to the ring geometry. This onsite splitting accounts for the waveguide shape of the structure. Formally, this splitting can be introduced in the tight-binding model a Hamiltonian term of the form:

$$H_{LT} = \Delta E \sum_{j=1}^{M} \left[ (\hat{\mathbf{a}}_j^\dagger \cdot \mathbf{e}_L^{(j)})(\mathbf{e}_L^{(j)\dagger} \cdot \hat{\mathbf{a}}_j) - (\hat{\mathbf{a}}_j^\dagger \cdot \mathbf{e}_T^{(j)})(\mathbf{e}_T^{(j)\dagger} \cdot \hat{\mathbf{a}}_j) \right]$$ (27)

with $\Delta E = \hbar\omega_R - \hbar\omega_A$. Expressing it in terms of the $k$-space operators, one gets a term of the form

$$H_{AR} = \frac{\Delta E}{2} \sum_k [\hat{a}_{k,\sigma_-}^\dagger \hat{a}_{k,\sigma_+} e^{-4\pi i/M} + \text{h.c.}]$$ (28)

with exactly the same form as the $\Delta t$ correction in (21). Equation (28) thus takes the form:

$$H = -\sum_{j=1}^{M} \{ \frac{\hbar t}{2} [\hat{\mathbf{a}}_{j+1}^\dagger \hat{\mathbf{a}}_j + \hat{\mathbf{a}}_j^\dagger \hat{\mathbf{a}}_{j+1}] +$$

$$-\sum_k \left( \frac{\hbar \Delta t}{2} \cos(\frac{\pi k}{3}) + \frac{\Delta E}{2} \right) [\hat{a}_{k,\sigma_-}^\dagger \hat{a}_{k,\sigma_+} e^{-4\pi i/M} + \text{h.c.}].$$ (29)

As the result, the eigenstates maintain the same form with the replacement

$$\Delta t \cos(\frac{\pi k}{3}) \to \Delta t \cos(\frac{\pi k}{3}) - \Delta_{AR}/2.$$ (30)

The lower $A$ eigenstate of the $k = 0$ manifold remains azimuthal polarized as long as $\hbar \Delta t > \Delta_{AR}/2$. The situation is slightly different for the $k = 3$ manifold, where the radially polarized state keeps a lower energy as long as $\Delta t > -\Delta_{AR}/2$. as sketched in the insets of Fig. 1f-g of the main text. The relative magnitude of the two effects has to be determined case by case on each specific structure. For instance, if instead of a hexagonal chain we would have a uniform ring guide, $\Delta_{AR}$ would be the dominant contribution to the SO coupling.

## II. SPIN-ORBIT HAMILTONIAN IN MATRIX AND OPERATOR FORM

We can gain insights on the emergence of the spin-orbit coupling from the polarisation dependent tunneling and onsite splittings by doing a matricial treatment of the problem. We again consider the basis of single pillar states with polarisations oriented longitudinal ($\mathbf{e}_L$) and transverse ($\mathbf{e}_T$) to the link between $j$ and $j+1$: $|j, L/T\rangle$. The polarisation dependent tunneling is described by single polariton Hamiltonian matrix elements:

$$\begin{aligned} t_L &= \left\langle j, L \left| \hat{H} \right| j+1, L \right\rangle \\ t_T &= \left\langle j, T \left| \hat{H} \right| j+1, T \right\rangle \end{aligned} \quad (31)$$

while

$$\left\langle j, L \left| \hat{H} \right| j+1, T \right\rangle = \left\langle j, L \left| \hat{H} \right| j+1, L \right\rangle = 0. \quad (32)$$

In order to include the onsite splitting between modes polarised in the direction radial and azimuthal to the ring geometry of the molecule, it is convenient to change to the polarisation basis $n, \tau$ depicted in Fig. 1:

$$\begin{aligned} |j, n\rangle &= \frac{\sqrt{3}}{2} |j, T\rangle - \frac{1}{2} |j, L\rangle \\ |j, t\rangle &= \frac{1}{2} |j, T\rangle - \frac{\sqrt{3}}{2} |j, L\rangle \\ |j+1, n\rangle &= \frac{\sqrt{3}}{2} |j+1, T\rangle - \frac{1}{2} |j+1, L\rangle \\ |j+1, t\rangle &= -\frac{1}{2} |j+1, T\rangle + \frac{\sqrt{3}}{2} |j+1, L\rangle . \end{aligned} \quad (33)$$

In this basis, the tight binding Hamiltonian reads:

$$\hat{H} = \begin{pmatrix} & 1,n & 1,\tau & 2,n & 2,\tau & 3,n & 3,\tau & 4,n & 4,\tau & 5,n & 5,\tau & 6,n & 6,\tau \\ 1,n & E_n & 0 & -t_{nn} & t_{n\tau} & 0 & 0 & 0 & 0 & 0 & 0 & -t_{nn} & -t_{n\tau} \\ 1,\tau & 0 & E_\tau & -t_{n\tau} & -t_{\tau\tau} & 0 & 0 & 0 & 0 & 0 & 0 & t_{n\tau} & -t_{\tau\tau} \\ 2,n & -t_{nn} & -t_{n\tau} & E_n & 0 & -t_{nn} & t_{n\tau} & 0 & 0 & 0 & 0 & 0 & 0 \\ 2,\tau & t_{n\tau} & -t_{\tau\tau} & 0 & E_\tau & -t_{n\tau} & -t_{\tau\tau} & 0 & 0 & 0 & 0 & 0 & 0 \\ 3,n & 0 & 0 & -t_{nn} & -t_{n\tau} & E_n & 0 & -t_{nn} & -t_{n\tau} & 0 & 0 & 0 & 0 \\ 3,\tau & 0 & 0 & t_{n\tau} & -t_{\tau\tau} & 0 & E_\tau & -t_{n\tau} & -t_{\tau\tau} & 0 & 0 & 0 & 0 \\ 4,n & 0 & 0 & 0 & 0 & -t_{nn} & -t_{n\tau} & E_n & 0 & -t_{nn} & -t_{n\tau} & 0 & 0 \\ 4,\tau & 0 & 0 & 0 & 0 & t_{n\tau} & -t_{\tau\tau} & 0 & E_\tau & -t_{n\tau} & -t_{\tau\tau} & 0 & 0 \\ 5,n & 0 & 0 & 0 & 0 & 0 & 0 & -t_{nn} & -t_{n\tau} & E_n & 0 & -t_{nn} & t_{n\tau} \\ 5,\tau & 0 & 0 & 0 & 0 & 0 & 0 & t_{n\tau} & -t_{\tau\tau} & 0 & E_\tau & -t_{n\tau} & -t_{\tau\tau} \\ 6,n & -t_{nn} & t_{n\tau} & 0 & 0 & 0 & 0 & 0 & 0 & -t_{nn} & -t_{n\tau} & E_n & 0 \\ 6,\tau & -t_{n\tau} & -t_{\tau\tau} & 0 & 0 & 0 & 0 & 0 & 0 & t_{n\tau} & -t_{\tau\tau} & 0 & E_\tau \end{pmatrix}, \quad (34)$$

where the onsite energy splitting $\Delta E = E_n - E_\tau$, and

$$\begin{aligned} t_{nn} &= \frac{1}{4}(3t_T - t_L) \\ t_{\tau\tau} &= \frac{1}{4}(3t_L - t_T) \\ t_{n\tau} &= \frac{\sqrt{3}}{4}(t_T + t_L). \end{aligned} \quad (35)$$

In order to find the explicit spin-orbit coupling terms of Hamiltoinan matrice2, it is convenient to change to the circular polarization basis $((|+\rangle, |-\rangle))$ instead of radial-azimuthal $((|n\rangle, |\tau\rangle))$ via the transformation:

$$\begin{aligned} |j, n\rangle &= \frac{1}{\sqrt{2}} \left[ exp(-i\pi \frac{j-1}{3}) |j, +\rangle + exp(i\pi \frac{j-1}{3}) |j, -\rangle \right] \\ |j, t\rangle &= \frac{1}{\sqrt{2}} \left[ exp(-i\pi (\frac{j-1}{3} + \frac{1}{2})) |j, +\rangle + exp(i\pi (\frac{j-1}{3} + \frac{1}{2})) |j, -\rangle \right] . \end{aligned} \quad (36)$$

Next, for the spatial component of the wavefunction, we change to the basis of orbital angular momentum $|l\rangle$ ($l = 0, \pm 1, \pm 2, 3$) via the transformation

$$|l, \pm\rangle = \sum_{j=1}^{6} exp(i\frac{l(j-1)\pi}{3}) |j, \pm\rangle. \tag{37}$$

In the orbital-circular polarisation basis, the Hamiltonian takes the form:

$$\begin{pmatrix} E-2J & 0 & 0 & 0 & 0 & 0 & 0 & -\frac{\Delta E}{2}-\frac{\hbar\Delta t}{2} & 0 & 0 & 0 & 0 \\ 0 & E-2J & 0 & 0 & 0 & 0 & 0 & 0 & -\frac{\Delta E}{2}-\frac{\hbar\Delta t}{2} & 0 & 0 & 0 \\ 0 & 0 & E-J & 0 & 0 & 0 & 0 & 0 & 0 & 0 & 0 & -\frac{\Delta E}{2}+\frac{\hbar\Delta t}{2} \\ 0 & 0 & 0 & E-J & -\frac{\Delta E}{2}-\hbar\Delta t & 0 & 0 & 0 & 0 & 0 & 0 & 0 \\ 0 & 0 & 0 & -\frac{\Delta E}{2}-\hbar\Delta t & E-J & 0 & 0 & 0 & 0 & 0 & 0 & 0 \\ 0 & 0 & 0 & 0 & 0 & E-J & 0 & 0 & 0 & -\frac{\Delta E}{2}+\frac{\hbar\Delta t}{2} & 0 & 0 \\ 0 & 0 & 0 & 0 & 0 & 0 & E+J & 0 & 0 & -\frac{\Delta E}{2}+\hbar\Delta t & 0 & 0 \\ -\frac{\Delta E}{2}-\frac{\hbar\Delta t}{2} & 0 & 0 & 0 & 0 & 0 & 0 & E+J & 0 & 0 & 0 & 0 \\ 0 & -\frac{\Delta E}{2}-\frac{\hbar\Delta t}{2} & 0 & 0 & 0 & 0 & 0 & 0 & E+J & 0 & 0 & 0 \\ 0 & 0 & 0 & 0 & 0 & -\frac{\Delta E}{2}+\frac{\hbar\Delta t}{2} & 0 & 0 & 0 & E+J & 0 & 0 \\ 0 & 0 & 0 & 0 & 0 & -\frac{\Delta E}{2}+\frac{\hbar\Delta t}{2} & 0 & 0 & 0 & 0 & E+2J & 0 \\ 0 & 0 & -\frac{\Delta E}{2}+\frac{\hbar\Delta t}{2} & 0 & 0 & 0 & 0 & 0 & 0 & 0 & 0 & E+2J \end{pmatrix} \begin{matrix} 0^+ \\ 0^- \\ +1^+ \\ +1^- \\ -1^+ \\ -1^- \\ +2^+ \\ +2^- \\ -2^+ \\ -2^- \\ 3^+ \\ 3^- \end{matrix}, \tag{38}$$

where

$$\begin{aligned} E &= \frac{1}{2}(E_\tau + E_n) \\ t &= \frac{1}{2}(t_L + t_T) \end{aligned} \tag{39}$$

This matricial form of the Hamiltonian is equivalent to Hamiltonian (29), showing the coupling between states of opposite orbital momentum and spin.

This Hamiltonian can also be expressed in operator form acting on a spinor $[\Psi_+(j)\Psi_-(j)]^T$, where $j$ plays a role of generalized integer coordinate, $j = 1,...,6$. For this we introduce the diagonal part of the Hamiltonian $\hat{H}_0 = \hat{H}(\Delta_{AR} = \hbar t = 0)$. The eigenstates of $\hat{H}_0$ can be classified in terms of the orbital angular momentum $l$ and produce the basis of Hamiltonian (38). Its eigenvalues $E_l$ are

$$E_l = E - 2Jcos\left(\frac{\pi l}{3}\right) \tag{40}$$

We can introduce an operator $\hat{M} = \frac{\partial^2 E_l}{\partial l^2} = cos\left(\frac{\pi l}{3}\right)$, which allows us to rewrite the Hamiltonian in the operator form:

$$\hat{H} = \hat{H}_0 - \frac{\Delta E}{2}\begin{pmatrix} 0 & e^{-2i\varphi_j} \\ e^{2i\varphi_j} & 0 \end{pmatrix} + \frac{\hbar\Delta t}{2}(\hat{M}\begin{pmatrix} 0 & e^{-2i\varphi_j} \\ e^{2i\varphi_j} & 0 \end{pmatrix} + \begin{pmatrix} 0 & e^{-2i\varphi_j} \\ e^{2i\varphi_j} & 0 \end{pmatrix}\hat{M}), \tag{41}$$

where $\varphi_j = j\pi/3$.

It is also possible to represent the same Hamiltonian in a more compact form using an operator $\hat{K}$ which returns the cosine of the sum of orbital momentum and spin:

$$\left\langle l\left|\hat{K}\right|l\right\rangle = cos(l + \sigma). \tag{42}$$

Then, Hamiltonian (38) can be expressed

$$\hat{H} = \hat{H}_0 - \frac{\Delta E}{2}\begin{pmatrix} 0 & e^{-2i\varphi_j} \\ e^{2i\varphi_j} & 0 \end{pmatrix} + \frac{\hbar\Delta t}{2}\hat{K}\begin{pmatrix} 0 & e^{-2i\varphi_j} \\ e^{2i\varphi_j} & 0 \end{pmatrix}. \tag{43}$$